\documentclass[twocolumn]{aastex62}
\usepackage{amsmath,enumitem}
\usepackage{lineno}

\newcommand{\srcfull}{FIRST~J141918.9+394036}
\newcommand{\src}{J1419+3940}

\usepackage{multirow}

\shortauthors{Mooley et al.}
\shorttitle{Follow-up of \srcfull}

\begin{document}

\title{Late-Time Evolution and Modeling of the Off-Axis Gamma-ray Burst Candidate \srcfull}


\author[0000-0002-2557-5180]{K.~P.\ Mooley}
\affil{National Radio Astronomy Observatory, Socorro, New Mexico 87801, USA}
\affiliation{Cahill Center for Astronomy and Astrophysics, MC 249-17 California Institute of Technology, Pasadena, CA 91125, USA}

\author[0000-0001-8405-2649]{B.~Margalit}
\altaffiliation{NASA Einstein Fellow}
\affiliation{Astronomy Department and Theoretical Astrophysics Center, University of California, Berkeley, Berkeley, CA 94720, USA}

\author[0000-0002-4119-9963]{C.~J.~Law}
\affiliation{Cahill Center for Astronomy and Astrophysics, MC 249-17 California Institute of Technology, Pasadena, CA 91125, USA}
\affiliation{Owens Valley Radio Observatory, California Institute of Technology, 100 Leighton Lane, Big Pine, CA, 93513, USA}

\author[0000-0001-8472-1996]{D.~A.\ Perley}
\affil{Astrophysics Research Institute, Liverpool John Moores University, IC2, Liverpool Science Park, 146 Brownlow Hill, Liverpool L3 5RF, UK}

\author{A.~T.~Deller}
\affiliation{Centre for Astrophysics and Supercomputing, Swinburne University of Technology, Hawthorn, Victoria, Australia}
\affiliation{ARC Centre of Excellence for Gravitational Wave Discovery (OzGrav), Australia}

\author{T.~J.~W.~Lazio}
\affiliation{Jet Propulsion Laboratory, California Institute of Technology, M/S 67-201, 4800 Oak Grove Dr., Pasadena, CA  91109}

\author{M.F.~Bietenholz}
\affiliation{Department of Physics and Astronomy, York University, Toronto, M3J 1P3, Ontario, Canada}
\affiliation{Hartebeesthoek Radio Astronomy Observatory, PO Box 443, Krugersdorp, 1740, South Africa}

\author{T.~Shimwell}
\affiliation{ASTRON, the Netherlands Institute for Radio Astronomy, Postbus 2, 7990AA, Dwingeloo, The Netherlands}
\affiliation{Leiden  Observatory,  Leiden  University,  PO  Box  9513,  NL-2300, Leiden, The Netherlands}

\author[0000-0002-5880-2730]{H.~T.\ Intema}
\affil{International Centre for Radio Astronomy Research, Curtin University, GPO Box U1987, Perth WA 6845, Australia}

\author[0000-0002-3382-9558]{B.~M.~Gaensler}
\affiliation{Dunlap Institute for Astronomy and Astrophysics, University of Toronto, 50 St. George Street, Toronto, Ontario, M5S 3H4, Canada}
\affiliation{Department of Astronomy and Astrophysics, University of Toronto, 50 St. George Street, Toronto, Ontario, M5S 3H4, Canada}

\author[0000-0002-4670-7509]{B.~D.~Metzger}
\affiliation{Department of Physics and Columbia Astrophysics Laboratory, Columbia University, New York, NY 10027, USA}
\affiliation{Center for Computational Astrophysics, Flatiron Institute, 162 W. 5th Avenue, New York, NY 10011, USA}

\author{D.Z.~Dong}
\affiliation{Cahill Center for Astronomy and Astrophysics, MC 249-17 California Institute of Technology, Pasadena, CA 91125, USA}

\author{G.~Hallinan}
\affiliation{Cahill Center for Astronomy and Astrophysics, MC 249-17 California Institute of Technology, Pasadena, CA 91125, USA}

\author{E.O.~Ofek}
\affiliation{Department of Particle Physics and Astrophysics, Weizmann Institute of Science, 76100 Rehovot, Israel}
    
\author{L.~Sironi}
\affiliation{Department of Astronomy, Columbia University, New York, NY 10027, USA}


\begin{abstract}
We present new radio and optical data, including very long baseline interferometry, as well as archival data analysis, for the luminous decades-long radio transient \srcfull.
The radio data reveal a synchrotron self-absorption peak around 0.3 GHz and a radius of around 1.3\,mas (0.5\,pc) 26 years post-discovery, indicating a blastwave energy $\sim5 \times 10^{50}$\,erg.
The optical spectrum shows a broad [OIII]$\lambda$4959,5007 emission-line that may indicate collisional-excitation in the host galaxy, but its association with the transient cannot be ruled out.
The properties of the host galaxy are suggestive of a massive stellar progenitor that formed at low metallicity.
Based on the radio light curve, blastwave velocity, energetics, nature of the host galaxy and transient rates we find that the properties of \src\ are most consistent with long gamma-ray burst (LGRB) afterglows. 
Other classes of (optically-discovered) stellar explosions as well as neutron star mergers are disfavored, and invoking any exotic scenario may not be necessary.
It is therefore likely that \src\ is an off-axis LGRB afterglow (as suggested by \citeauthor{2018ApJ...866L..22L} and \citeauthor{EVN}), and under this premise the inverse beaming fraction is found to be $f_b^{-1}\simeq280^{+700}_{-200}$, corresponding to an average jet half-opening angle $<\theta_j>\simeq5^{+4}_{-2}$ degrees (68\% confidence), consistent with previous estimates.
From the volumetric rate we predict that surveys with the VLA, ASKAP and MeerKAT will find a handful of \src-like events over the coming years.
\end{abstract}



\newpage
\section{Introduction}

The study of astrophysical transients is growing rapidly through a combination of new instruments, observing strategies, and theoretical advances. Extragalactic transients, such as supernovae (SNe), gamma-ray bursts (GRBs), and tidal disruption events (TDEs), are especially luminous and are typically produced during stellar death. New classes of extragalactic transient continue to be recognized \citep{2019NatAs...3..697I}.

High-energy and optical telescopes have traditionally dominated the discovery of energetic transients \citep{2004ApJ...611.1005G,2019PASP..131a8002B}. However, radio measurements are emerging as a valuable platform for transient discovery because radio wavelengths are sensitive to shocks formed by fast ejecta that may be expelled in such events \citep{1982ApJ...259..302C,1997ApJ...476..232M, 2001ApJ...562L..55F}. Radio synchrotron emission formed in shocks has a luminosity that is proportional to the total kinetic energy \citep{frail2005,2015MNRAS.454.3311M}. This fact has motivated a new generation of radio telescopes and surveys designed to be sensitive to transients \citep[e.g.,][]{2013PASA...30....6M,2017A&A...598A.104S,fender2017thunderkat,2020PASP..132c5001L}.

Early efforts to search for extragalactic radio transients were limited by lack of sensitivity or sky coverage \citep{2002ApJ...576..923L,2006ApJ...639..331G,2010ApJ...719...45C, 2011ApJ...742...49T,2011MNRAS.412..634B,2013ApJ...768..165M,2015MNRAS.450.4221B}.
New surveys can robustly detect sources brighter than $\sim$1 mJy over ten thousand square degrees. At this scale, the surveys are sensitive to radio spectral luminosities of $L_\nu=10^{30}$\ erg s$^{-1}$\ Hz$^{-1}$ over a volume of $\sim$1 Gpc$^{3}$, and can discover GRBs, TDEs, and more \citep{2015MNRAS.454.3311M}. Other factors that have traditionally limited radio transient discovery are issues related to correlator software (specifically for the legacy Very Large Array, e.g., phase-center noise), source significance statistics and lack of supporting multiwavelength measurements \citep[e.g.,][]{2006ApJ...639..331G,2011ApJ...742...49T,2012ApJ...747...70F}. These issues have been relieved though improvements to radio software/data analysis pipelines and better integration with follow-up observing resources \citep{2016ApJ...818..105M,2019ApJ...870...25M,2020MNRAS.491..560D,pintaldi2021}.

\srcfull\ \citep[hereafter \src;][]{2017ApJ...846...44O,2018ApJ...866L..22L} is an example showing the potential for radio discovery of extragalactic transients. \cite{2018ApJ...866L..22L} identified the source to be bright in the VLA FIRST survey \citep{1995ApJ...450..559B} in 1993, undetected by
the VLA Sky Survey \citep[VLASS;][]{2020PASP..132c5001L} in 2017, and a sub-milliJansky radio source in archival data steadily declining in flux density between 2010 and 2018. The transient was associated with a host galaxy, SDSS J141918.80+394035.9, at $z=0.01957$ \citep{sdss-dr9},
implying a radio spectral luminosity\footnote{The redshift z=0.01957 corresponds to a luminosity distance of 88.6\,Mpc and angular diameter distance is 85.2\,Mpc using {\it Planck} cosmological parameters \citep{planck-cosmology-2018}. We use these values throughout this paper.} of at least $2\times10^{29}$\ erg s$^{-1}$\ Hz$^{-1}$. This makes \src\ more luminous and longer lived than most supernovae, including those associated with LGRBs \citep{2016ApJ...830...42C}. The volume over which this source could have been detected is small (out to $\sim100$~Mpc), which implies a volumetric rate that is in tension with many of the known radio transient populations. 
\src\ is luminous, nearby, and at least three decades old, which makes it either a highly fortunate discovery or a prototype of a class of transient not well probed by past radio surveys.

No $\gamma$-ray, X-ray or optical counterparts of \src\ have been found and multiple origin models have been proposed. The luminosity, time scale, and host galaxy are consistent with an afterglow of a LGRBs \citep{2002ApJ...576..923L}. If so, the explosion occurred around 1993 with a total energy $E_j\sim10^{51}$\,erg that is interacting with a density of $n\sim10$\,cm$^{-3}$ \citep{2018ApJ...866L..22L} of circum-burst medium (CSM). 
The radio evolution and lack of a gamma-ray counterpart suggests that the event was an off-axis LGRB \citep[also known as orphan afterglow; e.g.,][]{ghirlanda2014}, the first of its kind. 
Very long baseline interferometric (VLBI) observations measured an
expansion speed of $0.1c$\ that is consistent with that hypothesis \citep{2019ApJ...876L..14M}. \citet{lee2020} proposed that \src\ could be the sign of interaction between ejecta from a neutron star merger and its surrounding interstellar material \citep{2011Natur.478...82N}. This model can explain the early light curve shape and is more consistent with the high volumetric rate implied by \src.

Other models for \src\ include new classes of transient powered by central engines. High-cadence optical surveys have defined a new class of engine-driven transients, akin to the prototype AT2018cow. 
AT2018cow-like events are a subclass of Fast Blue Optical Transients \citep[FBOTs; e.g.,][]{2013ApJ...774...58D} having luminous radio emission. Radio observations of AT2018cow-like transients have shown synchrotron emission from mildly relativistic outflows \citep{ho2019-2018cow,ho2020-koala,margutti2019,coppejans2020}. 
Potentially related is another new class of radio transient hypothesized to be associated with newborn magnetars \citep{2015MNRAS.454.3311M, 2016MNRAS.461.1498M}. 
This 
class is potentially frequent \citep{2017MNRAS.464.3568P} and radio luminous, but difficult to identify due to their long evolution timescale \citep[][but also see \citealt{2017ApJ...846...44O}]{margalit_metzger2018}.
Magnetar engines for transients are especially interesting, because they may leave a magnetar remnant long after the supernova. If so, this could tie the events to millisecond transients such as Fast Radio Bursts or highly luminous off-nuclear radio sources \citep{2017ApJ...846...44O,2019ApJ...886...24L, 2020arXiv200102688E}.

Here, we present new radio (including VLBI) and optical observations of transient \src, and a detailed interpretation of the transient source nature. New VLA and LOFAR data define a quasi-simultaneous radio spectrum from 0.15--10 GHz in two epochs and extend the time baseline of measurements at 1.4 GHz to 26 years. We also describe Very Long Baseline Array (VLBA) observations and reprocessing of previous VLA and European VLBI Network (EVN) datasets (\S\ref{sec:observe}). These measurements allow new analysis of the synchrotron blast wave energetics and a comparison to the direct measure of the shock expansion measured through VLBI observations (\S\ref{sec:modeling}). An analysis of the host galaxy and comparison with the host galaxies of known transients is presented in \S\ref{sec:host}. The properties of the radio transient and the host galaxy together with the event rate support the initial interpretation from \cite{2018ApJ...866L..22L} that \src\ is likely associated 
an with off-axis LGRB.
\S\ref{sec:OIII} gives possible explanations of a broad emission feature observed in the optical spectrum of \src. 
We present a summary and discussion of results in \S\ref{sec:summary} and end with the conclusions in \S\ref{sec:conclusions}.

\section{Observations, Data Processing and Initial Analysis}\label{sec:observe}

\subsection{VLA}\label{sec:vla}
We carried out observations with the NSF's Karl G. Jansky Very Large Array (VLA; under project code 19A-393, PI: Law), on 2019 May 18. Standard 8-bit WIDAR correlator setups were used for the P (300--500 MHz), L (1--2 GHz) and S (2--4 GHz) bands and 3-bit setups for the C (4--8 GHz) and X (8--12 GHz) bands to obtain the full frequency coverage possible with the VLA  up to 12 GHz. 
3C48 and PKS J0118-2141
were used as the flux and phase calibrators, respectively. 
The data were processed using the NRAO CASA pipeline (default version for CASA 5.6.1) and each band was split into an independent measurement set using CASA {\tt split}. Each measurement set was then imaged using CASA {\tt tclean} with natural weighting, pixel sizes chosen so as to resolve the synthesized beam with  $\gtrsim 4$ pixels,
image sizes appropriate to cover the primary beam full-width at half-maximum (FWHM), and a CLEAN stopping threshold of $3\times$ the thermal noise (as estimated from the VLA exposure time calculator).

\begin{table}
\centering
\caption{Radio data used in this work}
\label{table_chandra_specpar}
\begin{center}
\begin{tabular}{lllc}
\hline
\hline
Epoch   & Freq. & Flux Dens. & Telescope, Survey/Project\\
(year) & (GHz) & (mJy) \\
\hline
1993.87 & 1.465    & $26\pm2$       & VLA, AB6860 \\
1994.31 & 0.325    & $<9$           & WSRT, WENSS \\
1994.63 & 1.4 & 18.80 $\pm$ 0.95 & VLA, FIRST\\
1994.63 & 1.36 & 18.44 $\pm$ 0.94 & VLA, FIRST\\
1994.63 & 1.44 & 19.21 $\pm$ 0.98 & VLA, FIRST\\
1995.32 & 1.40     & 16.10$\pm$0.60 & VLA, NVSS\\
2008.54 & 1.415 & 2.5$\pm$0.2 & WSRT, ATLAS-3D\\ 
2010.64 & 1.415 & 1.70 $\pm$ 0.34 & WSRT, ATLAS-3D\\
2011.29 & 0.15 & $<$10 &  GMRT, TGSS\\
2015.36 & 1.33 & 1.212 $\pm$ 0.096 & VLA, 15A-033\\
2015.36 & 2.74 & 0.682 $\pm$ 0.082 & VLA, 15A-033\\
2015.36 & 3.43 & 0.562 $\pm$ 0.103 & VLA, 15A-033\\
2015.36 & 0.15 & $<$0.880  & LOFAR, LoTSS\\
2018.8 & 1.6 & 0.620 $\pm$ 0.095 & EVN, RM015\\
2019.37 & 2.3 & 0.415 $\pm$ 0.093 & VLBA, BL266\\
2019.46 & 0.36 & 1.991 $\pm$ 0.399 & VLA, 19A-393\\
2019.46 & 1.52 & 0.645 $\pm$ 0.045 & VLA, 19A-393\\
2019.46 & 3 & 0.343 $\pm$ 0.021 & VLA, 19A-393\\
2019.46 & 5.5 & 0.176 $\pm$ 0.011 & VLA, 19A-393\\
2019.46 & 9 & 0.093 $\pm$ 0.008 & VLA, 19A-393\\
2019.66 & 0.15 & 0.847 $\pm$ 0.202 & LOFAR, LoTSS\\
\hline
\end{tabular}
\tablecomments{All data points, except epochs 1993.87, 1994.31, 1995.32 and 2008.54, are new or revised from \cite{2018ApJ...866L..22L}. Absolute flux scale uncertainties (VLA: 5\%,  WSRT: 10\%, EVN/VLBA: 15\%, LOFAR: $\sim$10\%) have been added in quadrature with the statistical uncertainties. 
The WSRT ATLAS-3D data point at mean epoch 2010.64 represents the statistical mean and standard deviation of several data points reported by \cite{2018ApJ...866L..22L} around this epoch.
The TGSS upper limit has been corrected from the one reported in \citeauthor{2018ApJ...866L..22L}}
\end{center}
\end{table}

\begin{table}
\centering
\caption{Radio source size measurements}
\label{tab:size}
\begin{center}
\begin{tabular}{ccccc}
\hline
\hline
 Model     & Size parameter & Size (VLBA)    & Size (EVN)\\
           &                & (mas)               & (mas) \\
\hline
Shell       & Outer diameter & $2.5^{+0.7}_{-1.2}$    & $3.6_{-1.7}^{+4.6}$\\
Gaussian    & FWHM         & $1.5^{+0.4}_{-0.8}$    & $2.3_{-1.1}^{+2.7}$ \\
\hline
\end{tabular}
\tablecomments{
The fitted source size in mas for the VLBA and EVN datasets, with 1$\sigma$ uncertainties. All values were determined by
fitting geometrical models directly to the visibilities by least-squares.
The uncertainties include the statistical component and a systematic one
derived by allowing for 10\% uncertainty in the amplitude calibration of the individual antennas.  In the case of the EVN, where the range of data weights is high, we used the square root of the data weights in the fitting, which improves convergence at the expense of a small loss of
statistical efficiency. At an angular diameter distance of 85\,Mpc, 1\,mas $=$ 0.41\,pc $=$ $1.3\times10^{16}$\,cm.
}
\end{center}
\end{table}

\subsection{VLBA}\label{sec:vlba}

We observed J1419+3940 with the VLBA
at 2.3~GHz on 2019 April~12 (project BL266).  All ten VLBA antennas
were used.  
A standard continuum observing mode was used, with
eight spectral windows (each of~32~MHz width), and the observations
were conducted in a phase-referenced manner with ICRF J141946.6+382148 
used as
the phase reference calibrator (angular separation of 1.3 degrees).  The
phase reference cycle consisted of alternating scans of duration
approximately 4~min.\ scans on the science target J1419+3940 and
approximately 45~s on the phase-reference calibrator.

The data were calibrated in two different manners as a consistency
check, first using the rPICARD pipeline \citep[][using
CASA version~5.5]{janssen2019} 
second using AIPS (version 31DEC19).  
Because of considerable radio frequency interference (RFI),
only approximately half of the total bandwidth of 256 MHz was suitable for use.

We made a image using Briggs-weighting in CASA and robust=0.5, which had image background rms values of 31\,$\mu$Jy~beam$^{-1}$ and synthesized beam of 5.9 mas $\times$ 3.1 mas at 12\arcdeg\ position angle.  J1419+3940 appeared largely unresolved in the image.  To get a more precise idea
of the source size, we fitted a single circular Gaussian
component directly to the visibilities by weighted least-squares using the 
AIPS task \texttt{OMFIT}. 
The best-fit Gaussian had a flux density of $415 \pm 75 \; \mu$Jy,
where the uncertainty includes an assumed 15\% uncertainty on the flux-density scale.  

For marginally-resolved sources, the source size can be significantly 
correlated with any residual antenna amplitude miscalibration.  
We therefore determined the uncertainty on the FWHM size 
by including two components, added in quadrature: first,
the statistical contribution from the fit and second, the scatter obtained from a small 
Monte-Carlo trial where the antenna gains were randomized by 10\%, and the 
resulting visibility data re-fitted.  
In this case the statistical component dominated.
The measurement suggests a best-fit FWHM size of 
1.5~mas, but a completely unresolved source is excluded only at
the $1.7\sigma$ level.
The size measurement can also be compared with the FWHM value of
$3.9\pm0.7$ mas \citep{2019ApJ...876L..14M}, obtained with the European VLBI Network (EVN).
In \S\ref{sec:evn} we carry out an independent analysis of that EVN data.

However, for the radio source we are interested in a outer radius. 
For a circular Gaussian model the formal outer radius is infinity, so the Gaussian FWHM itself is not a appropriate estimator of the radio source size.
For marginally resolved sources, the fit is only very weakly
dependent on the choice of model.  The outer diameters of more physically appropriate models are related to the
Gaussian FWHM as follows:
uniform disk 1.60$\times$FWHM, optically thin shell 1.81$\times$FWHM.
Motivated by SNe 
we consider an optically-thin, uniform spherical shell model where
the shell thickness 20\% of the outer radius.
\citep[see,][and discussion therein]{SN2014C_VLBI2}.

We found the  best-fit outer diameter of such a shell to be 2.5~mas (and again, a completely unresolved
source is excluded only at the $1.7\sigma$ level).
The detailed model fit results are given in Table~\ref{tab:size}.


\subsection{LOFAR/LoTSS}

Observations were carried out with the Low Frequency Array (LOFAR; \citealt{2013A&A...556A...2V}) on 2015-07-28 (P214+40) and 2019-04-13 (P213+37) and as part of the ongoing 120-168\,MHz LOFAR Two-metre Sky Survey (LoTSS; \citealt{2019A&A...622A...1S,2017A&A...598A.104S}). The observations were processed following the current standard imaging procedures as described by \citet{2020arXiv201108328T}. The data were first calibrated to remove direction independent effects \citep{2016ApJS..223....2V, 2016MNRAS.460.2385W, 2019A&A...622A...5D} with the PreFactor\footnote{https://github.com/lofar-astron/prefactor} pipeline that uses the packages Default Pre-Processing Pipeline (DPPP; \citealt{2018ascl.soft04003V}), AOFlagger (\citealt{2012A&A...539A..95O}) and the LOFAR Solution Tool  (LoSoTo; \citealt{2019A&A...622A...5D}). To remove the remaining severe ionospheric and beam model errors, the data were then calibrated using the direction dependent self-calibration pipeline DDF-pipeline\footnote{https://github.com/mhardcastle/ddf-pipeline} that uses kMS \citep{2014A&A...566A.127T, 2015MNRAS.449.2668S} to derive direction dependent calibration solutions and DDFacet to apply these whilst imaging \citep{2018A&A...611A..87T}. The flux scale of the final images was refined using the procedure outlined by \cite{2020arXiv201108294H}. At a FWHM resolution of 6$\arcsec$, the final images have background rms noise levels of  260~$\mu$Jy~beam$^{-1}$ 
and 80~$\mu$Jy~beam$^{-1}$ at the pointing centres of P214+40 and P213+37, respectively, where the large discrepancy in background rms levels is due to unusually poor conditions during the P214+40 observation.

\subsection{Keck/LRIS}
\label{sec:observe:keck}

\begin{figure*}[htp!]
    \centering
    \includegraphics[width=18cm]{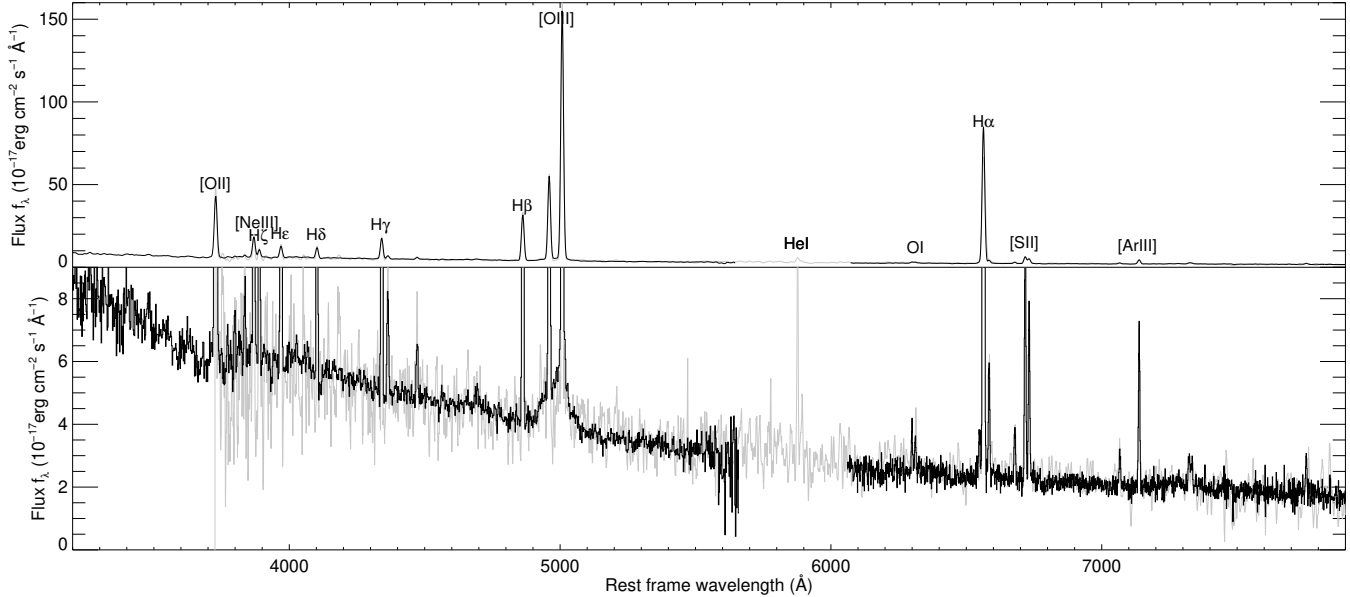}
    \caption{Spectroscopy of the host galaxy SDSS J141918.80+394035.9.  The SDSS spectrum (from 2004) is shown in gray and the Keck spectrum (from 2019) in black.  The upper and lower panels show different $y$-axis scalings of the data (spectra in the upper panel have also been convolved to a resolution of 12\AA).
    The host is an intensely starbursting galaxy.  A broad component is visible under the [OIII]$\lambda$4959,5007 line.}
    \label{fig:spectrum}
\end{figure*}
\begin{figure}[htp!]
    \centering
    \includegraphics[width=7cm]{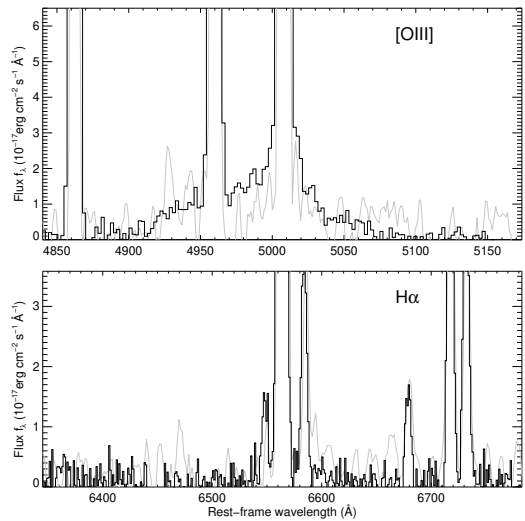}
    \caption{Zoom-in on [OIII] and H$\alpha$, with continuum subtracted.  The velocity scales and flux scales are the same way on each plot based on the central wavelength and integrated fluxes of the [OIII]$\lambda$5007 and H$\alpha$ lines, respectively.  The broad component underlying the [OIII] double is absent from H$\alpha$.}
    \label{fig:speczoom}
\end{figure}

The host galaxy of \src\ was observed with the Low-Resolution Imaging Spectrometer (LRIS; \citealt{1995PASP..107..375O}) at Keck Observatory on 2019-04-05 (PI: Hallinnan).  The blue-side spectrum (1$\times$1200 sec) was obtained using the 400/3400 grism and the red-side spectrum (2$\times$550 sec) was obtained using the 400/8500 grating.  Observations were reduced using LPIPE
\citep{2019PASP..131h4503P}.  Spectra were flux-calibrated using an observation of Feige 34 and the absolute scaling was adjusted to match an archival spectrum (taken in 2004) of the galaxy from the Sloan Digital Sky Survey (SDSS; also shown in Figure~\ref{fig:spectrum}).  The red-side CCD was incorrectly windowed during the observation, producing a small gap in the wavelength coverage.  

The reduced spectrum, shown in Figure \ref{fig:spectrum}, is that of a strongly star-forming, metal-poor galaxy (\S \ref{sec:host}).  However, a single broad feature is also evident underlying the narrow [OIII] lines.  If interpreted as a broad component of [OIII], the inferred velocity is $v \sim 3000$ km~s$^{-1}$, orders of magnitude in excess of the escape velocity of the low-mass host galaxy.  The origin of this feature is currently unclear.  No similar components are seen under any other narrow lines (or elsewhere in the spectrum). 

A zoom-in on the [OIII] and H$\alpha$ lines in shown in Figure~\ref{fig:speczoom}.
We estimate the broad component [OIII] flux to be about $5\times10^{-16}$ erg~cm$^{-2}$~s$^{-1}$, corresponding to a luminosity of $5\times10^{38}$ erg~s$^{-1}$, and a 3$\sigma$ upper limit of $1\times10^{38}$ erg~s$^{-1}$ for any broad H$\alpha$ component.
The SDSS spectrum also shows some hints of a broad [OIII] component, but it is difficult to ascertain its significance due to the larger noise and coarser resolution compared to the Keck spectrum.
Nevertheless, we find that if the broad component is present in 2004 then its luminosity is significantly lower than that measured in 2019.


\subsection{Reprocessing of Archival Data}

There are significant discrepancies in the radio spectral index and source size found between the results we present above and those previously reported in the literature \citep{2018ApJ...866L..22L,2019ApJ...876L..14M}.  In order to investigate potential sources of the discrepancies and to obtain improved estimates of the fitted values and/or uncertainties if possible, we reprocessed the archival data, and present the results below. 

\subsubsection{VLA/FIRST}
VLA  observations were performed on 1994 Aug 14th, 19th and 20th as part of the Faint Images of the Radio Sky (FIRST) survey \citep{1995ApJ...450..559B}, in two adjacent frequency bands centred on 1364.9 and 1435.1~MHz, with a bandwidth of 21.9~MHz in each band. Observations were calibrated in MIRIAD \citep{1995ASPC...77..433S}: data were flagged for radio frequency interference, calibrated for flux using observations of 3C~286, and had their time-dependent antenna gains estimated using observations of 3C~286 and QSO~B1504+377 (observed around an hour before and after the target field, respectively).
The calibrated UVFITS data were then imported into CASA.
Each epoch and each channel was independently imaged using {\tt clean} using pixel size of 1 arcsec, image size of 4000 pixels, Briggs weighting with the CASA robust parameter set to 0.5, single Taylor term and CLEAN stopping threshold in the range 0.7--0.9 mJy (roughly 4--5 times the thermal noise).
The three maps (each with a slightly different pointing center) for each frequency were then combined in the image plane using AIPS FLATN (with the appropriate primary beam correction defined in AIPS PBCOR).

\subsubsection{VLA/15A-033}
In order to verify the spectral index of the optically thin part of the radio spectrum at epoch 2015.36, we reprocessed the VLA/15A-033 (PI J. Farnes; galactic magnetic fields project) dataset, observed in the 1.5 GHz and 3 GHz bands.
The raw data were put through the NRAO pipeline built into CASA 5.6.1.
The processed data were then split into three frequency bins {\bf using CASA {\tt split}} and imaged using CASA {\tt clean} using a suitable pixel size to sample the synthesized beam with 4 pixels, image size suitable for imaging the FWHM primary beam, Briggs weighting with robust 0.5, two Taylor terms and CLEAN stopping threshold roughly $3\times$ the thermal noise.

\begin{figure*}
    \centering
    \includegraphics[width=\columnwidth]{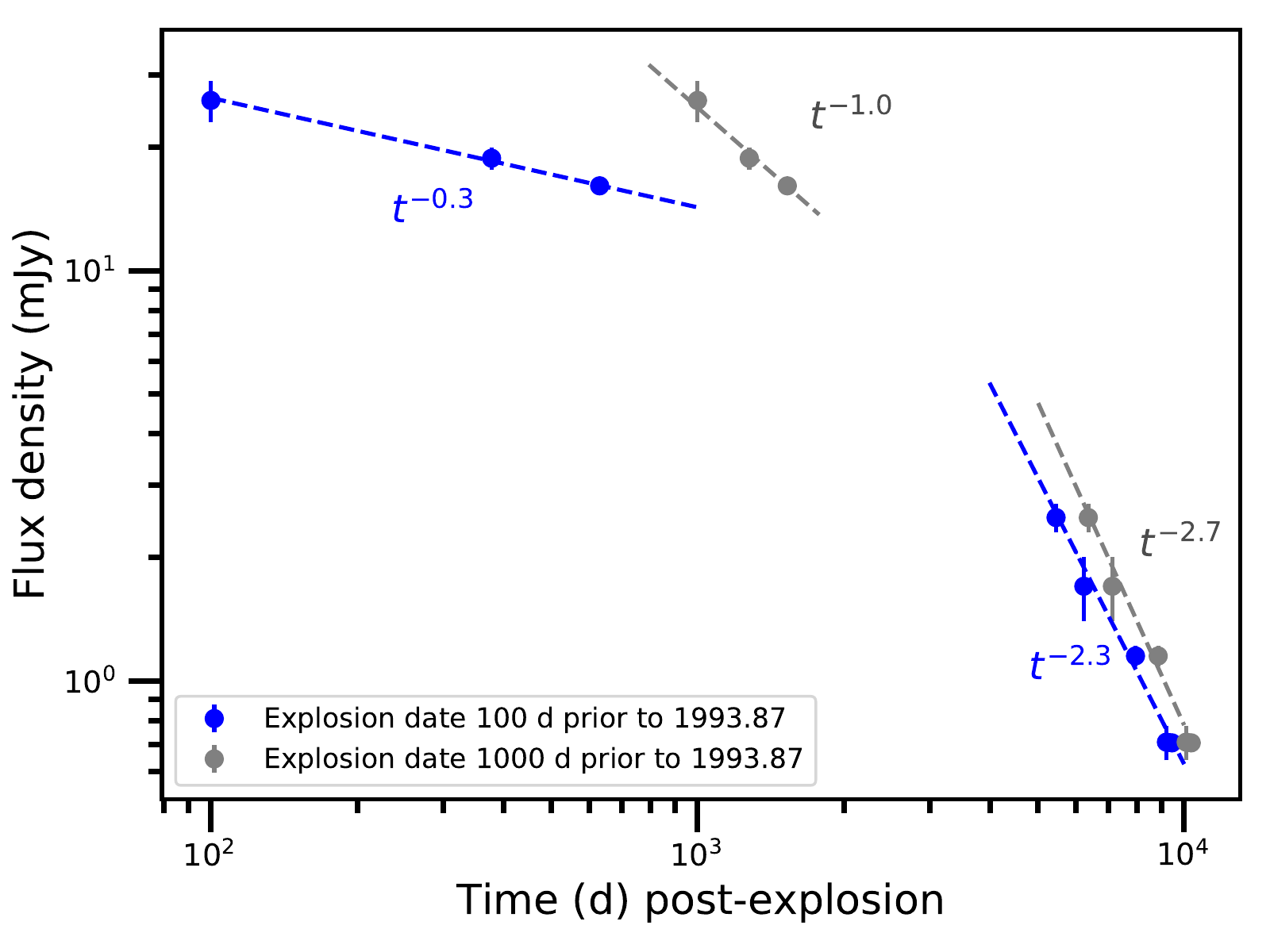}
    \includegraphics[width=3.5in]{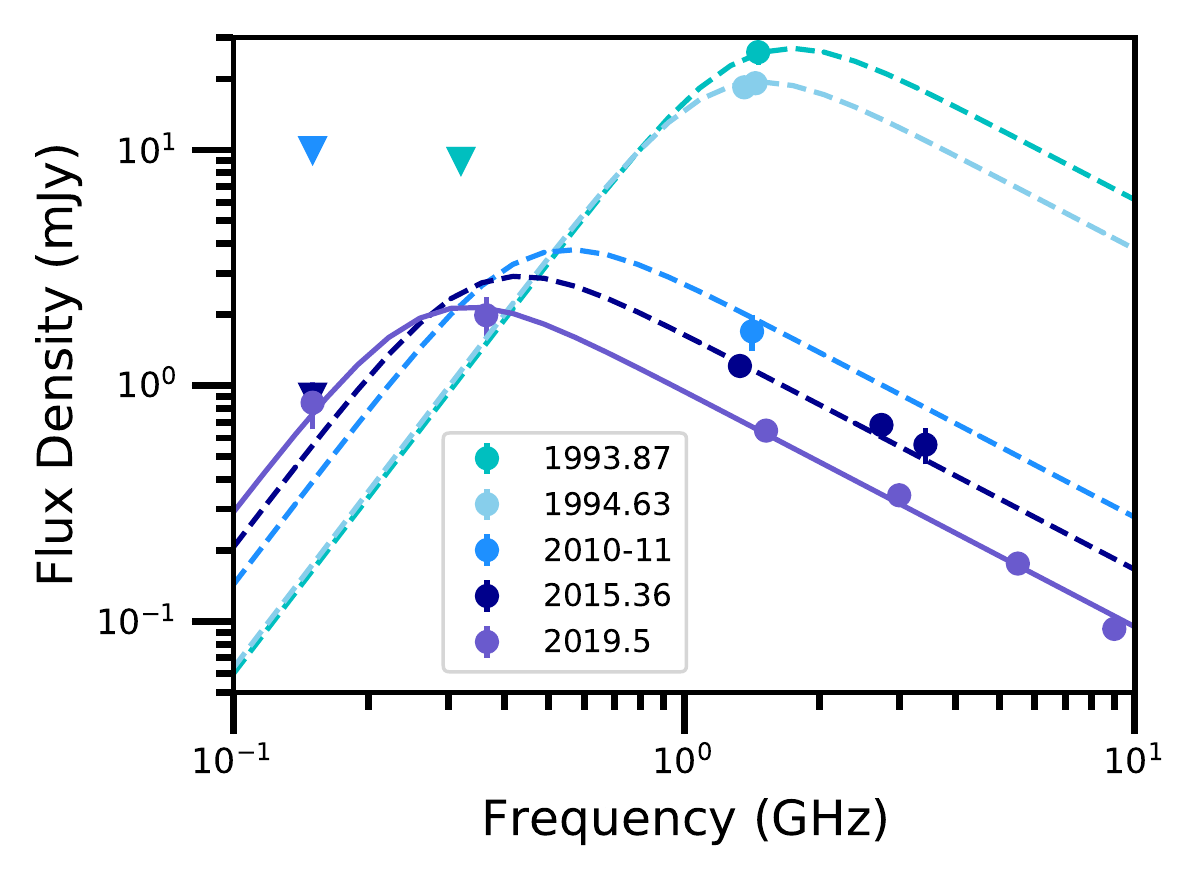}
    \caption{Left: The 1.4 GHz light curve of \src\ assuming two different ages at epoch 1993.87. The two colors, blue and gray, show the implied power-law for different assumed reference times of +100 d and +1000 d post-explosion, respectively. 
    Right: Radio spectral evolution for \src\ over five different epochs, spanning 26 years, from 1993.87 to 2019.5.
    The solid line is the best-fit smoothly-broken power-law model at epoch 2019.5. This model has been arbitrarily scaled in SSA frequency and peak flux density at different epochs, plotted as dashed curves, just to guide the eye (we do not use any of dashed curves in our analyses).}
    \label{fig:lcfit_sed}
\end{figure*}

\subsubsection{EVN}\label{sec:evn}
As described by \citet{2019ApJ...876L..14M}, EVN observations of \src\ were performed in September 2018 under project code RM015.  We downloaded the correlator data products from the EVN archive and reprocessed them using a ParselTongue pipeline \citep{Kettenis06}, which was adapted from that described by \citet{Mooley2018} with an additional step using the task \textit{APCAL} to load the {\em a priori} amplitude calibration corrections for EVN data.

We edited data during time periods 
affected by radio frequency interference, and for our final processing we also deleted all the data from the following stations: the Sardinia radio telescope (SRT), Cambridge, Deffin, and Knockin.  The SRT solutions displayed a high residual phase rate, while  the  other  three  telescopes  exhibited phase rate discontinuities. The SRT issue is thought to arise from a position error, while the discontinuities affecting the other three telescopes are believed to result from fibre delay corrections introduced by the WIDAR correlator (B.  Marcote,  priv.\  comm.)
However, we found that the inclusion or exclusion of these four antennas did not substantially bias the resulting size constraints.  
Regardless of this issue, the sparse $uv$ sampling of the dataset leads to challenges.
The longest baselines are primarily to just two stations: Tianma and Hartebeesthoek, and there are few baselines of intermediate length, meaning that (as noted by \citealp{2019ApJ...876L..14M}) the gain calibration for these two stations can considerably affect the fitted size.

After editing and calibration, we imaged the data using AIPS IMAGR (robust=0; uvtaper 40\,M$\lambda$) and found the fitted synthesized beam to be $5.50 \times 4.86$\,mas at a position angle of 71\arcdeg\ (the synthesized beam represents just the narrow inner lobe; sidelobes up to $\sim$80\% in amplitude are present across the broad plateau caused by the abundant shorter baselines).
The source appears to be marginally resolved in the image-plane, but we note that the negative extremum in the image is at 50\% the peak brightness, which is 379 $\mu$Jy~beam$^{-1}$ (RMS noise is 33 $\mu$Jy~beam$^{-1}$).  

We noted that the calibrated EVN data have a large range
of data weights, with a small fraction of the baselines, in particular those to a single
antenna, Effelsberg, having much higher weights than the remainder. 
To reduce the dominance of this small fraction of the
baselines, we used the
square-root of the original data weights for all baselines in the $uv$-plane modelfitting
\footnote{Using the square root of the weights serves to compress 
the range of weights, and thus reduces the dominance of a small number of high-weight baselines. This generally improves model-fitting convergence and
at the expense of a slight loss in statistical efficiency}.
Next, following a similar procedure adopted for the VLBA data (\S\ref{sec:vlba}), we fit both a spherical shell model and a Gaussian model in the $uv$-plane (using AIPS \texttt{OMFIT}) to find the source size.
\footnote{OMFIT uses $\chi^2$ minimization.  The $1\sigma$ uncertainties are determined by finding the points at which the $\chi^2$ increases over the best-fit value by a fraction of 1/(number of degrees of freedom).}
We find the best-fit values for the shell and Gaussian models in the $uv$-plane are: 1.8 mas (outer radius, corresponding to a source diameter of 3.6 mas) and 2.3 mas (FWHM) respectively, as given in Table~\ref{tab:size}.
These measurements are consistent, given the Gaussian FWHM to shell radius conversion factor of 0.8 (\S\ref{sec:vlba}).
However, we find that these nominal best-fit values have large uncertainties ($\sim$50--100\%, asymmetrical error bars), especially when compared with the relatively precise Gaussian FWHM measurement reported by \cite{2019ApJ...876L..14M} ($3.9\pm0.7$ mas; further discussed below).

In an attempt to reproduce the results of \citet{2019ApJ...876L..14M}, we
undertook a modelfitting procedure similar to the one 
used in that paper\footnote{Like \citet{2019ApJ...876L..14M} we used a $\chi^2$ technique. We sampled a dense grid of source positions, sizes, and peak amplitudes, recording the $\chi^2$ values at each point. Rather than relying on the absolute value of the $\chi^2$, we used the differential $\chi^2$ associated with a change in position from the best-fitting location to determine a confidence interval for the source size (using the positional uncertainty from an image-plane fit, which is relatively well-constrained).}, again fitting a circular Gaussian directly to the
visibilities by least squares, but using Difmap \citep{Shepherd97} rather
than AIPS \texttt{OMFIT}, and not taking the square-root of the nominal data
weights.  Using Difmap, we obtained a best-fit FWHM of 2.8 mas.  As
with \texttt{OMFIT}, we found the uncertainty range asymmetric, with a larger
uncertainty towards larger sizes.  Formally the Difmap $2\sigma$ range was $2.4 -
4.4$~mas, approximately consistent with the value of
$3.9 \pm 0.7$~mas published by \citet{2019ApJ...876L..14M}, but smaller than
the corresponding range we obtained using \texttt{OMFIT}.  The non-Gaussian image-plane
errors (with, as noted, the negative image extremum being at
$<-10\sigma$) strongly suggest that there are residual calibration
errors, which are probably not Gaussian-distributed, and being
antenna-based, would introduce correlations between the visibility
measurements.  Since the least-squares fits assumed that the errors in
the visibility measurements are Gaussian-distributed and independent,
this could be the cause of the discrepant uncertainty ranges between
\texttt{OMFIT} and Difmap.  In any case, given the low signal-to-noise ratio
and the likely presence of non-Gaussian-distributed errors,
we consider that the larger uncertainty range obtained from
\texttt{OMFIT} is likely more realistic for the size measurement from the EVN
data, and we use these results (as reported in Table~\ref{tab:size}) from this point onwards.

\begin{table*}
\centering
\caption{Source parameters derived from SSA analysis (following \cite{chevalier1998}).}
\label{tab:equipartition}
\begin{tabular}{llllll}
\hline
\hline
 & Epoch & $R$         & $B$  &   $U$        &     $n$\\
 &       & ($10^{17}$ cm) & (mG) & ($10^{49}$ erg) & (cm$^{-3}$) \\
 \hline
\multirow{2}{*}{$\epsilon_e=\epsilon_B=1/3$} & 1993.87 & 3.8 & 90 & 12  & 50$\times$(t$_{p,1.4\rm GHz}$/1000 d)$^2$\\
                                             & 2019.5  & 5.6 & 25 & 3 & 200 \\
\hline
\multirow{2}{*}{$\epsilon_e=0.1, \epsilon_B=0.01$} & 1993.87 & 3.3 & 60 & 100  & 800$\times$(t$_{p,1.4\rm GHz}$/1000 d)$^2$ \\
                                                   & 2019.5  & 5.0 & 15 & 25 & 2600 \\
\hline
\end{tabular}
\tablecomments{(1) For the 1993.87 epoch we have assumed that the SSA frequency is at 1.4 GHz. This assumption gives lower limits on R and U, upper limits on $B$ and $n$. (2) Uncertainties on the parameter values (based on the uncertainties on the fitted radio spectra) are approximately 10\%, 5\%, 5\% and 50\% for $R, B, U$ and $n$
respectively. The dependence on radius, $U\propto R^3$ and $n\propto R^2$, and the VLBI measurements (which imply $R\approx1.5\times10^{18}$\,cm in 2018/19) together indicate that the energy is underestimated by $\sim 25 \times$ and density is overestimated by $\sim10 \times$.}
\end{table*}

\begin{figure}[htp]
    \centering
    \includegraphics[width=\columnwidth]{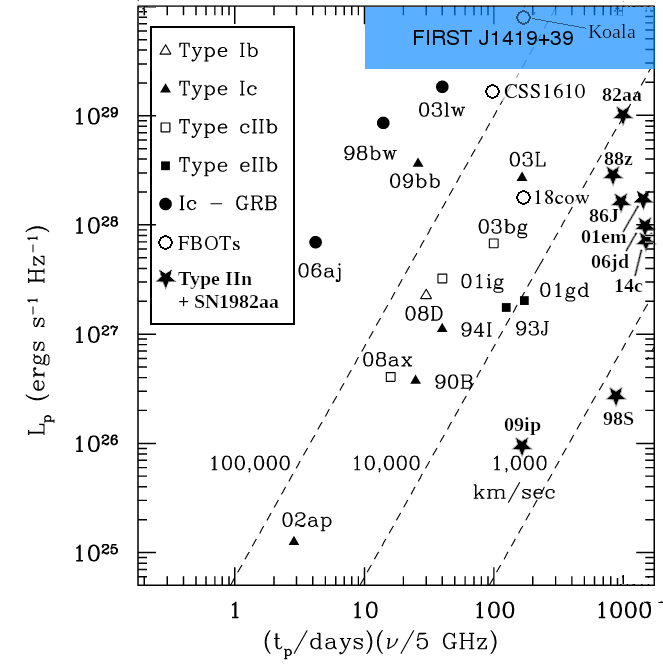}
    \caption{Luminosity versus timescale for optically-selected supernovae \citep[adapted from][]{chevalier_soderberg2010}. The blue region shows the part of the phase space occupied by \src. For \src\ and FBOTs, the peak luminosity and timescale around 1.4 GHz are used and for the rest of the sources the parameters are measured generally around 5--10 GHz. The dashed lines indicate the shock velocities as implied by the SSA peak frequency.}
    \label{fig:chevalier}
\end{figure}

\section{Modeling}\label{sec:modeling}

In this section we estimate the physical parameters based on the observational data. 
In view of the VLBI measurements presented in the previous section, we believe that the most appropriate measurement to use is the outer radius from the shell model. 
Considering the corresponding VLBA and EVN values listed in Table~\ref{tab:size}, we take the weighted mean to find the resulting outer radius of $1.3^{+0.3}_{-0.6}$ mas
\footnote{Although the associated uncertainty is large, the size measurement cannot immediately be dismissed as an upper limit since the source appears to be at least marginally resolved with the EVN. Further observations will be needed to improve the precision on this measurement. We also note that, since the measured size cannot be below zero, any measurement of the size will be biased high in the case of low SNR (like Ricean bias). For both EVN and VLBA, the best fit is about 2$\sigma$
above zero, so this bias may be significant and there will be a small upward bias in the weighted mean. Quantifying this bias is, however, non-trivial.}. 
This corresponds to a physical radius of $R = 0.5^{+0.1}_{-0.2} \, {\rm pc}$ at an angular diameter distance of 85 Mpc.

\subsection{Power-law fits to the radio light curve and spectra}
\label{sec:power_law_fits}
We fit the data points from early times (obtained in 1993--95) and late-time data points at $\sim$1.4~GHz with power-laws $F \propto t^b$. 
We do not know the time of explosion, so we try two fiducial values, 100 d and 1000 d \cite[motivated by the arguments presented by][]{2018ApJ...866L..22L}, for the age of the transient at the time of the first radio detection at epoch 1993.87. 
Assuming age 100 d (1000 d) post-explosion, we find $F_\nu \propto t^{-0.3}$ and $F_\nu \propto t^{-2.3}$ ($F_\nu \propto t^{-1.0}$ and $F_\nu \propto t^{-2.7}$; Figure~\ref{fig:lcfit_sed}) for the early and late-time respectively.

Simple power law fits, with $F \propto \nu^\alpha$ where $\alpha$ is
the spectral index, to the optically-thin spectra at epochs 2015.36 and 2019.46 give values of $\alpha$ of $-0.8\pm0.2$ and $-1.06\pm0.10$ respectively. 
The optically-thin spectral index is therefore consistent with $\alpha_{\rm thin}=-1$, suggesting that the electron power-law index is $p\approx3$ (where we have assumed that the GHz spectrum lies between the synchrotron self-absorption and cooling frequencies).
Fixing this spectral index and using $\alpha_{\rm thick}=+2.5$, we fit the epoch 2019.5 radio spectrum with a smoothly-broken power-law\footnote{Fitting a broken power-law, which corresponds a smoothly-broken power-law (SBPL) with smoothness parameter $s\xrightarrow{}\inf$, we get peak flux density $S_p=3.4\pm0.3$ mJy and the peak frequency 0.28$\pm$0.03 GHz. In this case, the parameter estimates given in \S\ref{sec:equipartition} change as $R\propto S_p^{9/19} \nu_p^{-1}$, $B\propto S_p^{-2/19} \nu_p$, $U\propto S_p^{23/19} \nu_p^{-1}$ compared with the smoothly-broken power-law case given in Table~\ref{tab:equipartition}. Specifically, $R$ increases by a factor of 1.3 compared to the SBPL case.} of the form described by \cite{beuermann1999,mooley2018-strongjet}, using MCMC to find\footnote{Since the turnover frequency depends on the flux densities of \src\ at 360 MHz and 150 MHz (obtained using different instruments), we verified the spectra of two nearby radio AGN, FIRST J141849.5+395154 and FIRST J141828.5+393928. 
Their spectra between 150 MHz, 360 MHz and 1.4 GHz appear perfectly consistent with a single power law with spectral index $-0.5$, indicating that there are no additional systematic offsets between the LOFAR and VLA P band data points.} the peak flux density, 2.25$\pm$0.52 mJy, the peak frequency, 0.30$\pm$0.04 GHz, and the smoothness parameter 0.39$_{-0.33}^{+0.45}$. The radio spectral evolution with these fits is shown in Figure~\ref{fig:lcfit_sed}.

\subsection{Spectral Evolution and Synchrotron Self-Absorption (SSA) Analysis}
\label{sec:equipartition}

We use the spectral parameters derived above to calculate the radius ($R$), magnetic field ($B$) and energy ($U$) using the \cite{chevalier1998} prescription \citep[see][for the relevant equations using $p\approx3$]{ho2019-2018cow}.
These calculated  values are tabulated
in Table~\ref{tab:equipartition}.
For the 1993.87 epoch we consider
for demonstrative purposes that the SSA frequency ($\nu_a$) is 1.4 GHz.
The average velocities implied by the equipartition ($\epsilon_e=\epsilon_B=1/3$) radii estimated at epochs 1993.87 and 2019.5 are about 44,000 ($t_{\rm d}/1000$ d)$^{-1}$ km~s$^{-1}$ and 7,000 ($t/26$ yr)$^{-1}$ km~s$^{-1}$ respectively (the former value is a lower limit since the peak luminosity at 1.4 GHz may be higher than that observed in 1993; $t_{\rm d}$ is the age of the transient at the discovery epoch 1993.87 and $t$ denotes the age around the VLBI observing epoch 2018/19).
In comparison, the average velocity 
as implied by the VLBI radius measurement of $\simeq1.2$ mas is about\footnote{The rest-frame time between the discovery epoch 1993.87 and the mean VLBI epoch 1995.1 is about 25 years, so we adopt a normalization of 26 yr for the age of the transient at the latter epoch.} 19,000 ($t/26$ yr)$^{-1}$ km~s$^{-1}$.
These velocities are reminiscent of Type Ib/c and Type Ic-broadline supernovae and make a Type II supernova explanation unlikely.

We also note that the cooling frequency $\nu_c$ is far above our observing band, and therefore irrelevant to this analysis. Specifically, 
from \cite{sari1998} we estimate the cooling frequency at epoch 2019.5 to be $\nu_c \sim $700\,GHz for the $\epsilon_e=\epsilon_B=1/3$ case (Table~\ref{tab:equipartition}) and even higher for lower values of $\epsilon_B$ ($\nu_c \propto \epsilon_B^{-3/2}$).


In Figure~\ref{fig:chevalier} we compare the peak luminosity and timescale of \src\ with those of different classes of stellar explosions, including FBOTs.
This figure also places the velocities derived above in the context of different radio afterglows, again indicating that a Type II supernova explanation is disfavored \citep[see also][]{bietenholz2021}.



\subsection{Energetics and light curve modeling}\label{sec:Ben}

Given the measured luminosity, spectrum, and radius of \src, we can place direct constraints on underlying properties of the source.
Typically in modeling synchrotron blast-waves there is a degeneracy between blast-wave energy and ambient-medium density. In the following the source size measurement can be used to break this degeneracy and unambiguously constrain the source energetics.

First, we explicitly define the formalism.
We assume that the observed radiation at $\gtrsim$GHz frequencies is produced by optically-thin synchrotron emission from a non-thermal population of electrons for which the momentum $\gamma \beta$ is distributed as a power-law $\propto \left(\gamma \beta \right)^{-p}$.
The optically-thin synchrotron spectrum is $\sim \nu^{-\frac{p-1}{2}}$ which, given the observed spectral index $\alpha \approx -1$, implies that $p \approx 3$ (see \S\ref{sec:power_law_fits}).
We further consider the standard scenario in which the synchrotron-emitting electrons are accelerated at a non-relativistic shock (with efficiency $\epsilon_e$), where magnetic fields are also amplified (with efficiency $\epsilon_B$).
For $p=3$, the synchrotron luminosity of such a blastwave 
can be expressed as (see Appendix~\ref{sec:Appendix_Synchrotron})
\begin{align}
\label{eq:nuLnu}
\nu L_\nu
\approx
10^{33} \, 
&{\rm erg \, s}^{-1}
\, \epsilon_{e,-1} \epsilon_{B,-1}
\left(\frac{t}{26 \, {\rm yr}}\right)^{-4}
\nonumber \\
&\times
\begin{cases}
4.8 
\left(\frac{R}{0.5 \, {\rm pc}}\right)^7  \left(\frac{m}{0.4}\right)^4 n_0^{2}
&;~{\rm ISM}
\\
8.9 
\left(\frac{R}{0.5 \, {\rm pc}}\right)^3 m^4 A_\star^{2}
&;~{\rm wind}
\end{cases}
,
\end{align}
where we have separated into two cases depending on whether the ambient medium
into which the shock expands
has a constant number density $n$ (ISM; normalized to $n_0 \equiv n / 1 \, {\rm cm}^{-3}$) or a wind-like density profile $\rho = A r^{-2}$ (wind; which we normalize to $A_\star \equiv A / 5 \times 10^{11} \, {\rm g \, cm}^{-1}$).
We have used the notation, $q_x = (q/10^x)$ in the appropriate unit for parameter $q$, e.g., $\epsilon_{B,-1} \equiv (\epsilon_B/0.1)$.
Equation~(\ref{eq:nuLnu}) is expressed in terms of the blastwave radius, $R$, and the source age $t$, both normalized to values consistent with \src.
The synchrotron luminosity is a strong function of velocity, the explicit dependence of which has been replaced above by taking
\begin{equation}
\label{eq:v}
v \equiv m R/t \simeq 
19,000 \,{\rm km \, s}^{-1}\,
\left(\frac{R}{0.5 {\rm pc}}\right) \left(\frac{t}{26 \, {\rm yr}}\right)^{-1} m
\end{equation}
such that $m \lesssim 1$ is a correction factor to the ``average'' velocity $R/t$ (in case of a power-law temporal evolution of the shock front, $m$ describes this exponent, i.e., $R \propto t^{m}$).
In the context of radio SNe, where the ambient medium is a wind environment, \cite{Chevalier82} shows that $m = (k-3)/(k-2)$ where $\rho_{\rm ej} \propto r^{-k}$ is the outer density profile of the SN ejecta, and $k \gtrsim 7$ are typical values (implying $m \gtrsim 0.8$).
On the other hand, a blastwave that propagates into a constant-density ISM and that is deep within the Sedov-Taylor regime will be characterized by $m=0.4$.
Finally, eq.~(\ref{eq:nuLnu}) has been derived assuming that the blastwave is in the so-called deep-Newtonian regime discussed by \cite{Sironi&Giannios13}. This regime is relevant if $v \lesssim v_{\rm DN} \simeq 0.15 c \, \epsilon_{e,-1}^{-1/2}$ (assuming $p=3$), and is therefore appropriate for \src\ at the current epoch (eq.~\ref{eq:v}).

Using eq.~(\ref{eq:nuLnu}) we can find the ambient density that is required in order to produce the observed spectral luminosity (at $1.5 \, {\rm GHz}$ at epoch\footnote{Since the spectral energy distribution goes as $L_\nu \sim \nu^{-1}$ above $\sim$1.5 GHz, $\nu L_\nu$ does not depend on frequency and hence any data point above 1.5 GHz would yield about the same result.} 2019.46) of \src, $\nu L_\nu \simeq 9 \times 10^{36} \, {\rm erg \, s}^{-1}$
It is
\begin{equation}
\label{eq:n_ISM}
n \approx 
43 \, {\rm cm}^{-3} \, \epsilon_{e,-1}^{-1/2} \epsilon_{B,-1}^{-1/2} \left(\frac{R}{0.5 \, {\rm pc}}\right)^{-7/2} \left(\frac{t}{26 \, {\rm yr}}\right)^{2} \left(\frac{m}{0.4}\right)^{-2}
\end{equation}
in the ISM case, and
\begin{equation}
\label{eq:A_constraint}
A_\star \approx 32 \, \epsilon_{e,-1}^{-1/2} \epsilon_{B,-1}^{-1/2} \left(\frac{R}{0.5 \, {\rm pc}}\right)^{-3/2} \left(\frac{t}{26 \, {\rm yr}}\right)^{2} m^{-2}
\end{equation}
for a wind medium.

Assuming that the blastwave in the ISM case is currently in the Sedov-Taylor regime (otherwise the light-curve would be rising rather than declining; e.g., \citealt{2011Natur.478...82N}), the energy associated with this blastwave is 
\begin{align}
\label{eq:E_ISM}
E &= \xi^{-5} n m_p R^5 t^{-2}
\nonumber \\
&\approx
4.6 \times 10^{50} \, {\rm erg} \, \epsilon_{e,-1}^{-1/2} \epsilon_{B,-1}^{-1/2} \left(\frac{R}{0.5 \, {\rm pc}}\right)^{3/2} , 
\end{align}
where $\xi \simeq 1.15$ is given by the Sedov-Taylor solution.
Note that eq.~(\ref{eq:E_ISM}) does not depend on the assumed source age. This energy is reasonable for various astrophysical sources, and in particular for LGRBs.

In the radio-SN case, \cite{Chevalier82} shows that the shock radius is
$R = \left[ 2 U_c^{k} / (k-4)(k-3) A \right]^{\frac{1}{k-2}} t^{\frac{k-3}{k-2}}$ 
where $U_c$ is a parameter governing the outer ejecta density profile, $\rho_{\rm ej} = \left( r / U_c t \right)^{-k} t^{-3}$. This parameter can be related to the ejecta energy that is contained above some velocity coordinate, $E_{\rm ej}(\geq v) = \frac{2 \pi}{k-5} U_c^{k} v^{-(k-5)}$. Using these expressions, we can estimate a lower limit on the ejecta energy that is required for interpreting \src\ within the radio-SN paradigm,
\begin{align}
\label{eq:Eej_radioSN}
E_{\rm ej}(\geq v) 
&= \frac{\pi (k-4)(k-3)}{(k-5) m^{k-5}} R^3 t^{-2} A
\nonumber \\
&\gtrsim 4 \times 10^{51} \, {\rm erg} \,
\epsilon_{e,-1}^{-1/2} \epsilon_{B,-1}^{-1/2} \left(\frac{R}{0.5 \, {\rm pc}}\right)^{3/2}
.
\end{align}
The second line is for $k \simeq 6.3$ (and correspondingly $m \simeq 0.77$), at which $E_{\rm ej}$ attains a minimum as a function of $k$ (this accounts for the $m=(k-3)/(k-2)$ dependence, and the additional $m$ dependence implied by eq.~\ref{eq:A_constraint}).
Different values of $k$ would imply larger ejecta energies. For example, for the physically-motivated value of $k \approx 12$ \citep{Matzner&McKee99}, we find that $E_{\rm ej} \sim 10^{52} \, {\rm erg}$.
Furthermore, eq.~(\ref{eq:Eej_radioSN}) only accounts for the energy of ejecta material that has velocity $v \gtrsim 20,000 \, {\rm km \, s}^{-1}$ 
(as inferred in eq.~\ref{eq:v}) so that the total ejecta energy (including lower-velocity material) is likely to be even higher.
This energy constraint disfavors the interpretation of \src\ as a typical\footnote{Note that we have here adopted the \cite{Sironi&Giannios13} framework for the deep-Newtonian regime, different from much of the radio-SN literature. Beyond the physical motivation for this approach, we note that it is conservative in the sense that it predicts higher luminosities at late times. Adopting the \cite{chevalier1998} approach would therefore imply even larger $E_{\rm ej}$.} radio-SN \citep[see also ][]{2017ApJ...846...44O}.

\begin{figure*}[htp!]
    \centering
    \includegraphics[width=\columnwidth]{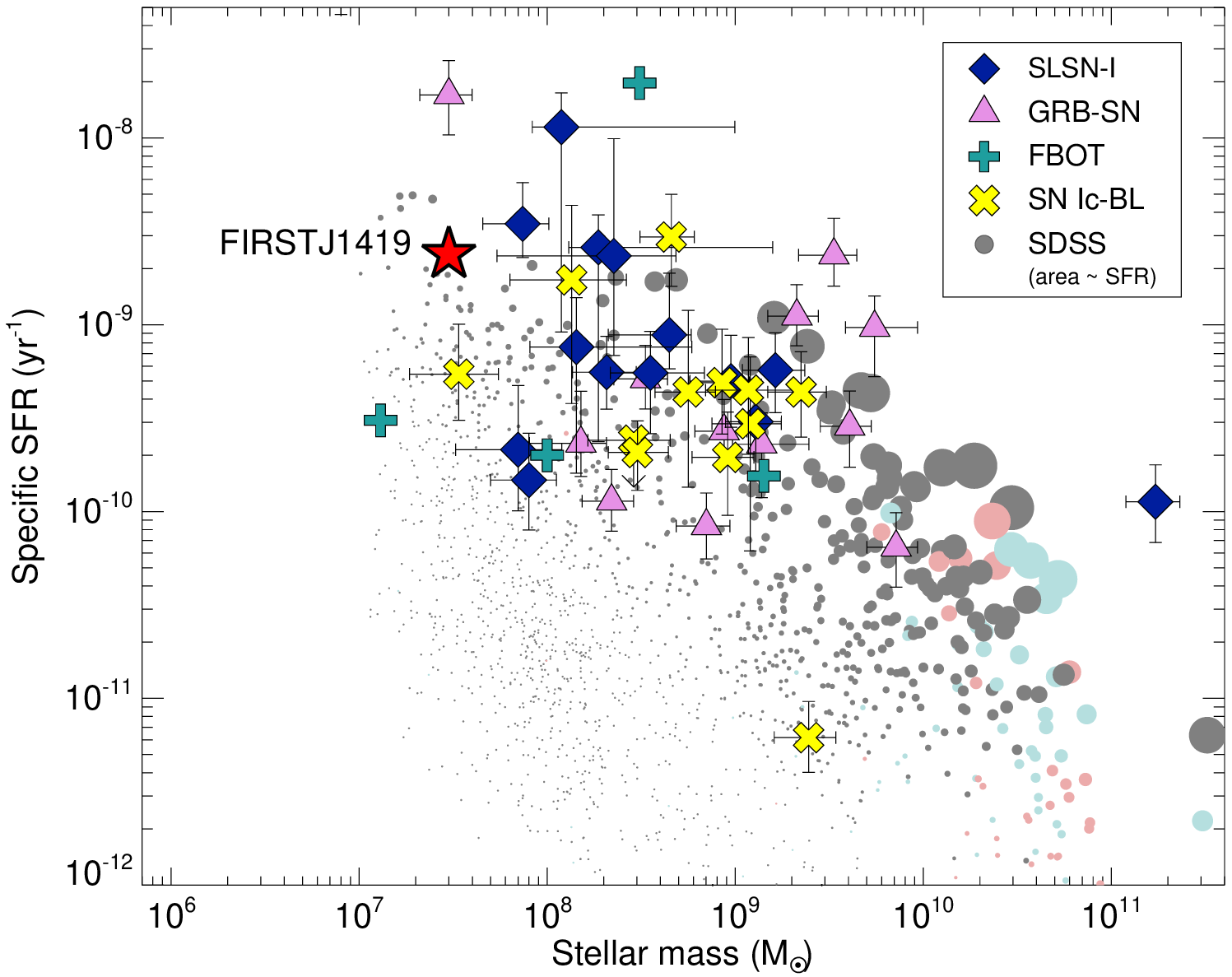}
    \includegraphics[width=\columnwidth]{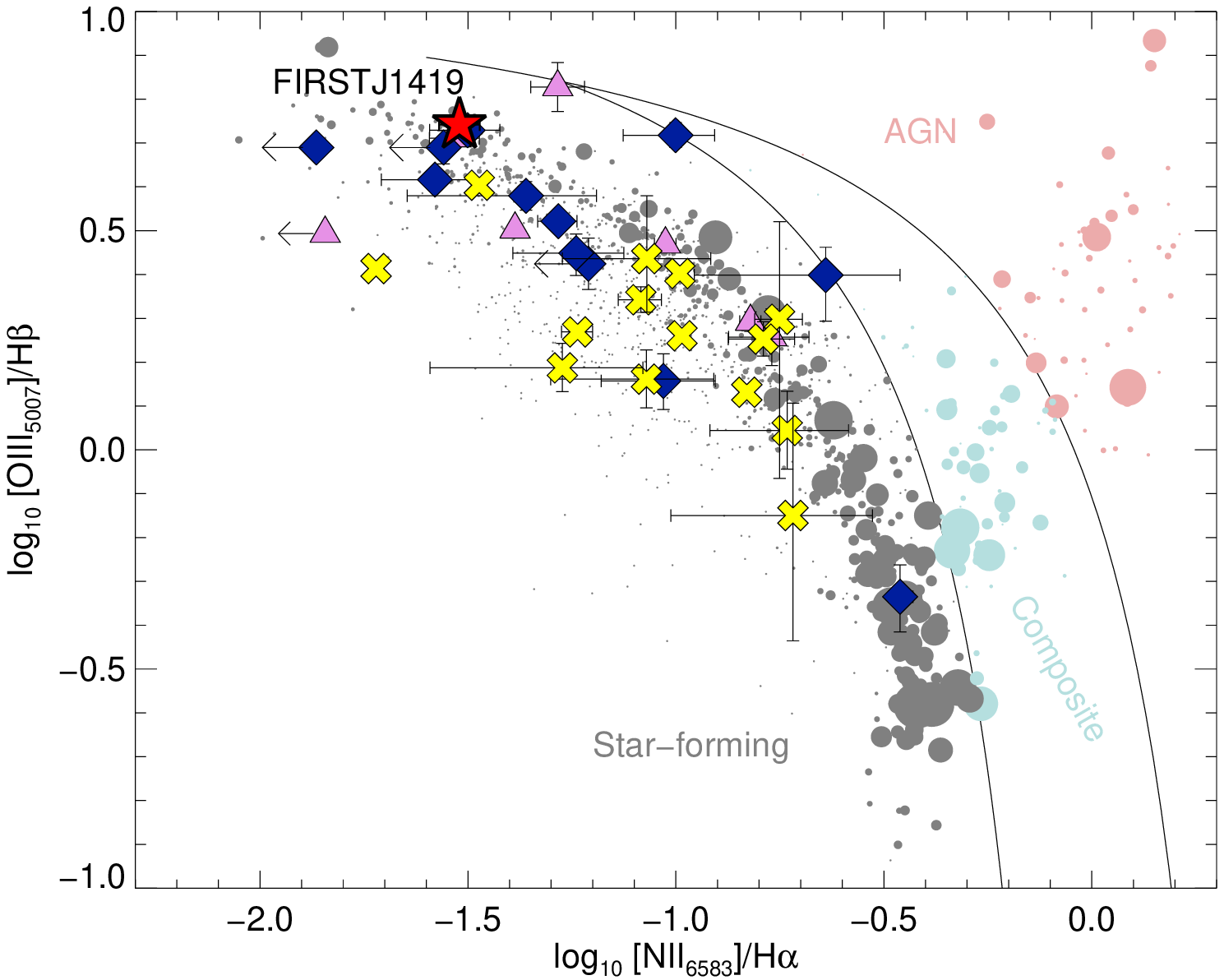}
    \caption{Left: Mass versus specific star-formation rate for the host galaxy of \src\ and the hosts of a variety of comparison samples: superluminous supernovae (blue diamonds), GRBs (purple triangles), Ic-BL SNe (yellow crosses), FBOTs (cyan pluses) and SDSS galaxies (gray circles).  SDSS galaxies have been resampled to simulate a volume-limited survey and the size of the points are scaled by SFR to visually represent a SFR-selected sample.  SDSS galaxies with AGN contamination are recolored.
    Right: BPT line-ratio diagram for the host galaxy of \src\ as compared to the same comparison samples as in the left panel.
    Solid lines show the AGN separation criterion.  The emission properties of the host galaxy of \src\ are typical of low-redshift star-forming dwarf galaxies.}
    \label{fig:host}
\end{figure*}


\subsubsection{Initial velocity}

Having ruled out the typical radio-SN scenario (and hence wind-medium) on energetic grounds, we hereon out focus on the constant-density (ISM) blastwave scenario.
As discussed above, the fact that \src's light-curve is observed to decline implies that the shock is within the Sedov-Taylor regime, $t > t_{\rm dec}$, 
i.e., the shock must be decelerating. This implies that the shock velocity at the current epoch (eq.~\ref{eq:v}) is lower than the initial blast-wave velocity $v_i$. We can roughly constrain this initial velocity by considering the deceleration (or Sedov-Taylor) timescale $t_{\rm dec}$
\citep[e.g][]{Hotokezaka&Piran15},
\begin{align}
\label{eq:t_dec}
t_{\rm dec} 
&= \left(\frac{3 E}{4\pi m_p c^5 n \Gamma_i^7 \left(\Gamma_i-1\right)\beta_i^3}\right)^{1/3}
\\ \nonumber
&\approx 
\begin{cases}
33 \, {\rm yr} \, E_{{\rm ej},51}^{1/3} n_0^{-1/3} \beta_{i,-1}^{-5/3}
&,~\beta_i \ll 1
\\
0.1 \, {\rm yr} \, E_{{\rm ej},51}^{1/3} n_0^{-1/3} \left(\Gamma_i/2\right)^{-8/3}
&,~\Gamma_i \gg 1
\end{cases}
.
\end{align}
The deceleration time depends on the {\it initial} blast-wave velocity $v_i$ (where $\beta_i = v_i/c$ and $\Gamma_i = 1/\sqrt{1-\beta_i^2}$ is the corresponding Lorentz factor), and we specifically consider also the possibility that the initial blast-wave was highly relativistic (bottom case). For a spherical explosion (and neglecting synchrotron self-absorption),
the light-curve rises up to $t \sim t_{\rm dec}$ and subsequently declines.

Equation~(\ref{eq:t_dec}) shows that the light-curve rise time is $\sim$decades for outflows whose initial velocity is sub-relativistic. 
In particular, $t_{\rm dec}$ would be $\gtrsim 16 \, {\rm yr}$ if the initial velocity $v_i$ were similar to the current inferred velocity (eq.~\ref{eq:v}). This is in tension with the data, that instead suggest a much faster light-curve rise time. The fact that
the observed flux density of \src\ declines by a factor of $\sim$2 in the span of $\approx 1.5$ yr (between epochs 1993.87 and 1995.3) since the first detection epoch
is suggestive of a rise timescale $\lesssim$2 years.
Using eqs.~(\ref{eq:n_ISM},\ref{eq:E_ISM},\ref{eq:t_dec}) we can use this as a constraint on the initial velocity. Requiring that $t_{\rm dec} \lesssim 2$ years implies that the initial velocity must have been
\begin{equation}
v_i \gtrsim 0.2 c \, \left(\frac{R}{0.5 \, {\rm pc}}\right) \left(\frac{t}{26 \, {\rm yr}}\right)^{-2/5} \left(\frac{m}{0.4}\right)^{2/5}
.
\end{equation}
at the very least trans-relativistic.
Another (albeit not mutually-exclusive) possibility is deviation from spherical symmetry, such as an initially collimated explosion pointed off-axis from our line-of-sight. The short rise time in this case may be attributed to a jet break, however this scenario would also require an initially relativistic outflow (for the case of non-relativistic outflows, asymmetry has only a modest effect; e.g., \citealt{Margalit&Piran15}).


\subsubsection{Synchrotron self-absorption}

As a final point, we can estimate the synchrotron self-absorption (SSA) frequency implied by the ISM-case solution for \src\ (eq.~\ref{eq:n_ISM}).
We estimate $\nu_{\rm SSA}$ by equating the optically-thin synchrotron luminosity (eq.~\ref{eq:nuLnu}) to the optically-thick luminosity $L_{\nu,{\rm SSA}} \sim 8\pi^2 m_e R^2 \gamma(\nu) \nu^2 /3$, where $\gamma(\nu) = (2\pi m_e c \nu / e B)^{1/2}$ and $B$ the magnetic field. This approximate approach is correct up to order-unity correction factors due to the, uncertain, geometry and the electron distribution function.
In this manner, we find that (for $p=3$)
\begin{align}
\nu_{\rm SSA} 
\approx 
&16 \, {\rm MHz} \, 
\epsilon_{e,-1}^{2/7} \epsilon_{B,-1}^{5/14} n_0^{9/14}
\\ \nonumber
&\times 
\left(\frac{R}{0.5 \, {\rm pc}}\right)^{11/7} \left(\frac{t}{26 \, {\rm yr}}\right)^{-9/7}
\left(\frac{m}{0.4}\right)^{9/7}
.
\end{align}
For the inferred ISM density of \src\ (eq.~\ref{eq:n_ISM}), this implies
\begin{equation}
\nu_{\rm SSA} \approx 170 \, {\rm MHz} \, \epsilon_{e,-1}^{-1/28} \epsilon_{B,-1}^{1/28} \left(\frac{R}{0.5 \, {\rm pc}}\right)^{-19/28} 
,
\end{equation}
which depends almost exclusively on the shock radius $R$.
This result is broadly consistent with the observed spectrum (at this epoch, $\sim$2019), which exhibits a turnover at a few hundred
MHz that is compatible with SSA (Fig.~\ref{fig:lcfit_sed}; see also \S\ref{sec:power_law_fits}).
Note that precise details of the transition between the optically-thick and optically-thin regime depend on additional geometric effects, which we do not consider here \citep[e.g][]{Bjornsson&Keshavarzi17}.

\section{Host Galaxy}\label{sec:host}

The host galaxy of the transient, SDSS J141918.80+394035.9, is a blue compact dwarf and is clearly detected in SDSS survey imaging \citep[see][]{2018ApJ...866L..22L}.   
The spectra are shown in Figure \ref{fig:spectrum}.  We extracted line fluxes by fitting a Gaussian profile to each emission line, and obtained estimates of the gas-phase oxygen abundance using both strong-line methods (all 13 diagnostics enabled within {\tt pymcz}, \citealt{Bianco+2015}) as well as a $T_e$-based ``direct'' measurement based on the [OIII]$\lambda$4363 auroral line \citep{Izotov+2005}.  Other bulk measurements (including mass and star-formation rate, SFR, estimates) were obtained from the NASA-SDSS Atlas (NSA; \citealt{2011AJ....142...31B}).  We infer a stellar mass of $M_* = 3 \times 10^{7} M_\odot$, specific SFR (sSFR) of $2.4\times10^{-9}$ yr$^{-1}$, and an oxygen abundance of 12+log[O/H] = 8.10$^{+0.06}_{-0.05}$ ($T_e$ method; strong-line methods give consistent estimates.)

The left panel of Figure~\ref{fig:host} shows the SED-derived stellar mass ($M_*$) and specific star-formation rate (SFR/$M_*$) of the host galaxy of \src\ in comparison to the SDSS spectroscopic sample and some other populations known to reside within extreme galaxies: SLSNe and GRBs \citep{2016ApJ...830...13P}, Ic-BL SNe \citep{Modjaz+2020} as well as a 1/$V_{\rm max}$ resampling of the NSA spectroscopic sample.  \src\ is at the extreme of all of these groups, but it is not an outlier.  Similarly, the right panel of Figure~\ref{fig:host} plots two strong-line emission ratios (as in \citealt{1981PASP...93....5B}); the host lies near one end of the diagram due to its low metallicity but is ionization properties are otherwise consistent with a normal star-forming dwarf.

While it is impossible to come to a secure conclusion about the nature of the progenitors of \src-like transients based on a single event, these properties derived above generally support a massive stellar origin and favor a progenitor that is intrinsically more likely to form at low mass, high sSFR, or low metallicity.  Gamma-ray bursts, Ic-BL supernovae, superluminous supernovae, and AT2018cow-like fast transients all seem to have these properties \citep{2016A&A...593A.115J,2020ApJ...892..153M,2018MNRAS.473.1258S,2021arXiv210301968P}.  An exotic origin is otherwise not required, as while the properties of the host are not typical they are far from unprecedented (approximately 1 in 50 supernovae explodes in a similarly extreme environment; \citealt{2019arXiv191109112T}).

\begin{figure*}[htp]
    \centering
    \includegraphics[width=6.5in]{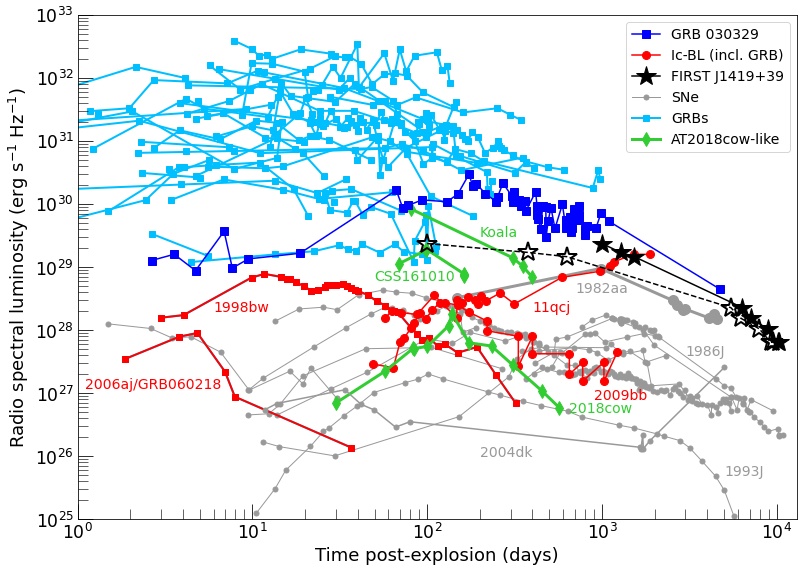}\\
    \caption{Light curve of \src\ (1.4 GHz) compared with GRBs (1--5 GHz), supernovae (1--5 GHz) and AT2018cow-like FBOTs (1--5 GHz, but 10 GHz for Koala). The black stars represent the light curves of \src\ (unfilled stars assume \src\ is 100 d old at epoch 1993.87 and filled stars assume an age of 1000 d at the same epoch). We note that the late-time light curve decline of \src\ is strikingly similar to that of GRB030329. The light curves are compiled from \cite{2012ApJ...746..156C} (GRBs), \cite{palliyaguru2021} (PTF 11qcj), \cite{soderberg2010} (SN 2009bb), \cite{balasubramanian2021,wellons2012} (SN 2004dk), \cite{ho2019-2018cow,ho2020-koala,coppejans2020} (FBOTs with luminous radio emission, i.e., AT2018cow-like events), 
    \cite{Palliyaguru+2019, weiler1986,  Weiler+1991, Montes+2000,
    yin1994, WeilerPS1990, SN86J-1, SN86J-3, SN93J-2, Chandra+2009a,
    Schinzel+2009, Stockdale+2004, SN2001em-1, SN2001em-2, soderberg2006,
    Soderberg+2006c, Chandra+2012a, salas2013, Margutti+2017_SN2014C, SN2014C_VLBI, bietenholz2021,  Ryder+ATel8836, Argo+ATel9147, Terreran+2019} (other SNe), 
    and unpublished processing of archival data.} 
    \label{fig:lc}
\end{figure*}

\begin{figure*}[htp]
    \centering
    \includegraphics[width=5.2in]{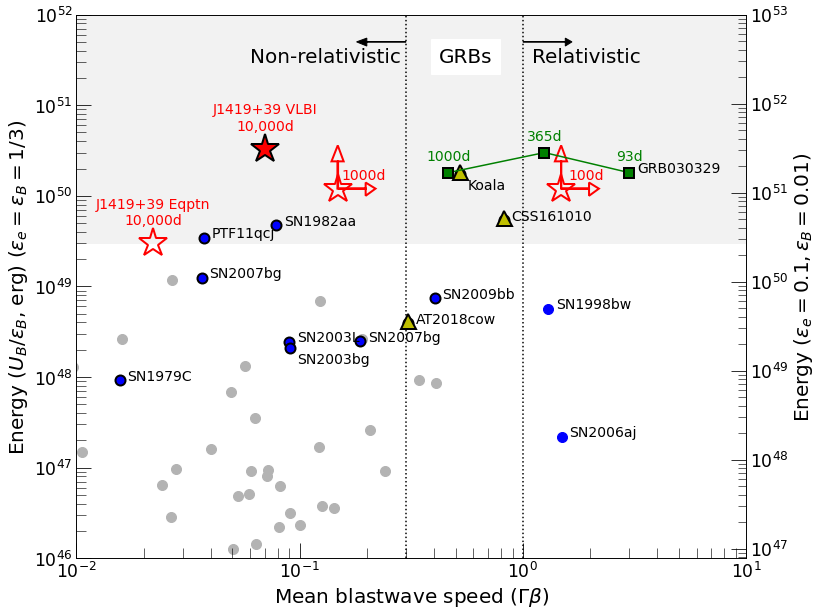}
    \caption{Energy versus mean blastwave speed for \src~(red stars) in comparison with other stellar explosions. Unfilled stars denote the positions based on the SSA analysis while the filled star denotes the position inferred from the VLBI measurement. 
    In this plot, the epoch at which the SSA peak and the source radius are measured (2018/19), is assumed 10,000~d post-explosion.  
    Blue circles denote SNe \citep{weiler1986,yin1994,kulkarni1998,soderberg2005,soderberg2006,soderberg2006-2003bg,soderberg2010,salas2013,palliyaguru2021} where the equipartition parameters are calculated at $\sim$50--1500 days post-explosion (except SN2006aj for which the 5 GHz peak is around 5 days), when the SSA peak lies between 1--5 GHz. Two data points plotted for SN2007bg represent the two peaks observed in its 6 GHz light curve. Grey circles are SNe from the compilation of \cite{bietenholz2021} where the SSA peak is taken to be the peak of the light curve (observing frequency between 5--10 GHz). 
    Yellow triangles denote AT2018cow-like events \citep{ho2019-2018cow,margutti2019,ho2020-koala,coppejans2020}. 
    GRBs occupy the phase space shown by the grey region. 
    Green squares denote the evolution of the afterglow of GRB030329 \citep{frail2005,vanderhorst2008}.}
    \label{fig:U_v}
\end{figure*}

\section{Origin of the broad [OIII]$\lambda$4959,5007 line}
\label{sec:OIII}

Here we consider several possibilities for the broad ($\sim$3000 km~s$^{-1}$) [OIII] line observed in the Keck/LRIS spectrum (\S\ref{sec:observe:keck}).

\begin{enumerate}[leftmargin=0.5cm]

\item[a)] The broad component could originate from collisional broadening in extremely dense star-forming regions.  Broad profiles of similar widths have been seen in some of the most extreme H\,{\sc ii}\ regions in the SMC \citep{1985A&A...145..170T,1999ApJ...518..246K} although to our knowledge it has not been reported in integrated spectra of entire galaxies.  The host galaxy is extremely star-forming and moderately metal-poor, so this interpretation is quite plausible.  
A possible challenge to this interpretation is the lack of a similar broad component to the H$\alpha$ line.

\item[b)] It could also originate from a fast outflow driven by star-formation.  However, the relatively high velocity makes this possibility unlikely (typical galactic winds are of order 100 km~s$^{-1}$, with even 1000 km~s$^{-1}$ considered to be extreme; \citealt{2005ARA&A..43..769V}).

\item[c)] An active galactic nucleus (AGN) is another possible explanation. 
Some dwarf galaxies are known to harbor AGN  
(or ``proto-AGN"; \citealt{mezcua2019,Halevi+19,reines2020}; but see \citealt{2020arXiv200102688E})
although it is unusual for an AGN to show broad [OIII] and no broad H$\alpha$.

\item[d)] Finally, it is possible that the broad component may be coming from the transient itself. 
Broad [OIII] due to CSM-interaction has been observed in decades-old supernovae, but in all cases luminous hydrogen emission is also produced \citep[e.g.,][]{milisavljevic2012}. Nebular spectra of hydrogen-poor SLSNe also show hydrogen features (although nebular spectra decades post-explosion have not been reported) \citep[e.g.,][]{yan2015,nicholl2016}. 
Thus SN-CSM interaction cannot\footnote{Whether SN-CSM interaction with an unusually H-poor environment \citep[see e.g.,][for brief discussions]{chatzopoulos_wheeler2012, 2018ApJ...864L..36M} can explain the spectrum remains to be explored.} explain the spectrum of \src.
Broad nebular emission lacking hydrogen lines years after a supernova has been seen in at least one previous case \citep{2018ApJ...864L..36M}, although only on timescales of a few years and not decades, and in that case other oxygen lines were observed that are not apparent here.
\src\ occurred prior to the SDSS spectrum shown in Figure \ref{fig:host} being taken, and while (given the lower sensitivity) it is not completely clear whether the line is \emph{absent} in 2004, it certainly was not any brighter in comparison to 2019, which makes this explanation relatively unlikely.

\end{enumerate}

Given these possibilities we consider a collisional broadening component to be the most conservative interpretation, but the other possibilities cannot be entirely ruled out.  Some discussion of the possible implications if the [OIII] line can be attributed to the transient are discussed in \S\ref{sec:summary}.

\section{Summary \& Discussion}\label{sec:summary}

\subsection{Summary of radio observations and derived spectral information}
We have carried out late-time radio (VLA 1--12 GHz, LOFAR 0.15 GHz and VLBA 3 GHz) and optical spectroscopic follow up observations, and reprocessing of archival radio data of the transient \srcfull~
\citep[\src;][]{2018ApJ...866L..22L}.
For the first time we unambiguously determine the peak in the afterglow spectrum around 0.3 GHz (at epoch 2019.46, i.e. 26 years after the first detection of the transient with the VLA at epoch 1993.87). 
We identify the peak with the synchrotron self-absorption (SSA) frequency and the optically-thin part of the spectrum lying between the SSA and cooling frequencies.
The optically-thin part of the spectrum at epoch 2019.46 satisfies $F_\nu \propto \nu^{-1}$ (Figure~\ref{fig:lcfit_sed}), indicating a electron power-law index of $p\simeq3$.
This radio spectral index of $\alpha_{1-10~\rm GHz}\simeq-1$ can be compared with the X-ray-to-radio constraint of $\alpha_{R-X}<-0.25$ implied by the {\it Swift} X-ray flux upper limit \citep{2018ApJ...866L..22L}. 
The late-time decline of the 1.4 GHz light curve is consistent with $t^{-2.3}$ ($t^{-2.7}$) assuming that the first detection occurred 100 d (1000 d) post-explosion. 
These fits, shown in Figure~\ref{fig:lcfit_sed}, do not suggest any further steepening of the light curve around epoch 2017/18 \citep[suggested earlier by][based on the 3 GHz non-detection in the VLASS Epoch 1 quick-look data and the 1.5 GHz EVN flux density]{2018ApJ...866L..22L,2019ApJ...876L..14M}.

\subsection{Decline of the late-time radio light curve is relatively fast}
The inferred light-curve decline rate is steeper than that expected ($\sim t^{-1.2}$) for a subrelativistic Sedov-Taylor blastwave within the deep-Newtonian regime \citep{Sironi&Giannios13}, however this is inferred over a small dynamical range in time ($\lesssim 0.3\,$dex) and is sensitive to the assumed explosion epoch.
We note that a decline rate $\sim t^{-2.4}$, more consistent with the observed light-curve, is expected if the minimal Lorentz factor for the population of radio-emitting electrons is $\gg 1$ \cite[e.g.,][]{2000ApJ...537..191F}. However \cite{Sironi&Giannios13} show that this is only possible if the shock velocity exceeds $\simeq 0.15 c \, \epsilon_{e,-1}^{-1/2}$, which is not satisfied by \src\ (eq.~\ref{eq:v}). This velocity threshold could be lowered if only a small fraction $\zeta_e \ll 1$ of electrons that are swept-up by the shock participate in diffusive-shock acceleration, as long as the total energy carried by these electrons remains significant (large $\epsilon_e$). If the decline-rate of \src\ is interpreted this way, this would imply $\zeta_e < 0.01 \, \epsilon_{e,-1} \left(R/0.5\,{\rm pc}\right)^2 \left(t/26\,{\rm yr}\right)^{-2}$ and constitute a novel constraint on this parameter (typically implicitly taken to be $\zeta_e=1$).
Finally, we note that a steeply declining light-curve could point instead towards a drop in the CSM density profile encountered by the shock and/or changing microphysical parameters (time varying $\epsilon_e$, $\epsilon_B$), though it is unclear whether these scenarios are well-motivated in the case of \src.

\subsection{Summary of VLBI results and parameters derived from the radio analysis}

The source radius, measured using VLBA and EVN (data obtained in 2018/19), is measured to be $1.3^{+0.3}_{-0.6}$ mas, i.e. $R = 0.5^{+0.1}_{-0.2}$ pc for an angular-diameter distance of 85 Mpc.
This implies an average velocity of 19,000 km\,s$^{-1}$ ($\simeq 0.06 c$) over the
$\sim$26 years evolution of \src\ (see eq. \ref{eq:v}).
This size constraint allows us to constrain the blastwave energy and ambient density of \src\ independently. 
For a constant density (e.g., ISM) circumstellar medium, we find $n \approx 40 \, {\rm cm}^{-3}$ and $E \approx 5 \times 10^{50}$ erg (eqs.~\ref{eq:n_ISM},\ref{eq:E_ISM}). 
These values also imply a SSA peak at $\sim$170 MHz.
These results are broadly consistent with an SSA analysis (\S\ref{sec:equipartition}, Figure~\ref{fig:lcfit_sed}), which suggests a blastwave energy $10^{49}-10^{51}$ erg, and magnetic field strength $\sim$10--100 mG.

\subsection{Summary of optical spectroscopic findings}

In the optical we find a broad, $\sim$3000 km\,s$^{-1}$, [OIII]$\lambda$4959,5007 line that may be collisionally-excited nebular emission from compact star-forming region(s) in the host galaxy, but we cannot confidently rule out its association with \src.

\subsection{Classification of the transient}

No multiwavelength counterparts have been detected for \src\ apart from the emission features described above. Using our radio observations we can estimate the late-time properties of the blastwave, but it is difficult to ascertain the progenitor and whether the transient was initially relativistic.
However, we consider the following lines of argument to further investigate the nature of the transient.
Given the properties of the host galaxy (\S\ref{sec:host}), we primarily consider stellar explosion scenarios. 

\subsubsection{Peak spectral luminosity, light curve evolution, comparison with the afterglows of past events.}
In Figure~\ref{fig:lc} we compare the light curves of GRBs and SNe with \src. The peak luminosity of $>2\times10^{29}$ erg\,s$^{-1}$\,Hz$^{-1}$ at 1.4 GHz is unprecedented for regular SN afterglows \citep{weiler2002,bietenholz2021}, but it is compatible with GRB afterglows (or their associated SNe Ic-BL, e.g., SN 1998bw) and possibly AT2018cow-like events.
Acknowledging the caveat that a comparison between the afterglows of optically-selected SNe and the radio-selected afterglow of \src\ could be biased, we proceed with the following discussion.

In rare cases dense-CSM interaction, as seen in SN IIn, Ibn and Ic-BL, can produce high radio luminosity. 
Especially in the case of SN Ic-BL PTF 11qcj \citep{palliyaguru2021}, the afterglow is currently undergoing rebrightening due to late-time CSM interaction (the 1.4 GHz luminosity is currently about $10^{29}$ erg\,s$^{-1}$\,Hz$^{-1}$ at $\simeq$3000 days post-explosion; see Figure~\ref{fig:lc}).
We cannot immediately rule out CSM interaction based on the properties of the afterglow light curve, but we return to this point below.

In Figures~\ref{fig:chevalier} and~\ref{fig:lc} we have shown well-known SN Ic-BL: 98bw, 03lw, 06aj (GRB-associated), 09bb, 02ap and 11qcj (not GRB-associated).
It is evident that \src\ is much more luminous ($\sim$2x--20,000x) and longer-lived than normal radio-loud Ic-BL (but not rising to late-times like the interacting 11qcj, discussed above). 
Moreover, most SN Ic-BL are not radio-detected \citep[e.g.][]{2016ApJ...830...42C}.
Hence, a simple Ic-BL afterglow explanation for \src\ appears to be unlikely (but  CSM-interacting or off-axis GRB-associated Ic-BL remains plausible).

We can compare \src~with AT2018cow-like events \citep{ho2019-2018cow,ho2020-koala,margutti2019,coppejans2020}, but since only a handful of such events are currently known, their properties remains uncertain and the comparison cannot be conclusive.
Although the peak luminosity of \src\ (at 1.4 GHz) is much larger than that observed for any of the SN2018cow-like events, one such event (CSS161010, \citealt{coppejans2020}) has a luminosity approaching $10^{29}$\,erg\,s$^{-1}$\,Hz$^{-1}$. 
However, the radio light curves (around 1.4 GHz) of such FBOTs are generally seen to peak on timescales of few 100 days and decline rapidly (faster than $t^{-3}$; e.g. \citealt{ho2020-koala}), unlike the $\sim t^{-2.5}$ and decades-long emission seen for \src.

The sample size of radio-detected SLSNe is even smaller \citep{2019ApJ...876L..10E,Eftekhari+21,2019ApJ...886...24L,coppejans2021}, and although their 1.4 GHz radio spectral luminosities are $<10^{28}$ erg\,s$^{-1}$\,Hz$^{-1}$ their association with \src\ cannot be immediately ruled out (but see below).

Only one previous radio-discovered afterglow, SN 1982aa \citep[Mrk 297A;][]{yin1994}, is known to have a peak spectral luminosity around $10^{29}$ erg\,s$^{-1}$\,Hz$^{-1}$ at 1.4 GHz and peak timescale of $\sim$1000 days. 
\cite{bietenholz2021} suggest, based on the luminosity and timescale, that SN1982aa may be a SN IIn, but the nature of the transient remains uncertain since no optical spectrum is recorded.
\src\ differs from SN1982aa in that the late-time decline in the radio light curve ($t^{-2.5}$) and the radio spectrum ($\nu^{-1}$) are much steeper than those of SN1982aa ($t^{-1.3}$ and $\nu^{-0.75}$).
The properties of SN1982aa are generally in agreement with those measured for radio SNe (\citealt{weiler2002,yin1994,bietenholz2021}; and notably, also similar to SN1998bw), while \src\ appears to be an outlier in this group. 

The peak luminosity, timescale and decline of the radio light curve of \src\ are similar to those seen for some GRB afterglows.
Long GRBs have peak 1.4 GHz spectral luminosities around $10^{30}$ erg\,s$^{-1}$\,Hz$^{-1}$ (short GRBs have lower peak spectral luminosities), peak timescale around $\sim$few--100 days, and the light curve decline at late times is $\sim t^{-1}- t^{-2}$.
For off-axis events the peak luminosity and timescale may be longer \citep[as in the case of the neutron star merger GW170817;][]{dobie2018,fong2019}, depending on the observing angle.
The late-time evolution of off-axis and on-axis GRB afterglows is expected to be similar \citep[e.g.,][]{granot2002,salafia2016,kathirgamaraju2016}.
Note that if we assume the age of \src\ to be $\sim$1000 d in 1993, then the light curve of \src\ is strikingly similar to that of GRB030329 (see Figure~\ref{fig:lc}).

Taken together, the light curve of \src\ is dissimilar to the afterglows of optically-selected supernovae (SN II, Ib/c and Ic-BL; i.e., not associated with jets / central engines) and to the handful of AT2018cow-like events that are currently known, but consistent with GRB afterglows.
The connection with SLSN-I cannot be ruled out purely based on light curve arguments, but we return to the case of SLSN-I below.


\subsubsection{Velocity.} 
As described in \S\ref{sec:modeling}, an average velocity of at least $\sim$44,000\,($t_{\rm d}$/1000 d) km\,s$^{-1}$ is needed to explain the first radio detection of \src\ in 1993, and if the rise in the light curve is as rapid as the decline, then initial velocity needs to be $\gtrsim$0.2c. 
The average velocity up to the mean VLBI observing epoch 2019.1 
is $0.06 c$.
Such velocities (together with the peak radio luminosity) can be reconciled only by SN Ic-BL or engine-driven explosions/jets and rule out SN II \citep[e.g.,][]{bietenholz2021}. 
These velocities are also compatible with the dynamical ejecta from neutron star mergers, as pointed out by \cite{lee2020}, and we return to this point below.


\subsubsection{Energetics.} 
In Figure~\ref{fig:lc}, we show blastwave energy as a function of velocity for \src\ and other stellar explosions.
The energy and velocities are large compared to most supernovae, but comparable to those of GRBs and AT2018cow-like events.
Specifically, as also shown in \S\ref{sec:Ben}, the energy required to explain the  properties of \src\ is very large for typical radio SNe, and makes this scenario unlikely. 
The energy reservoir of $10^{51}$ erg also makes SN 2009bb-like events \citep[][i.e., engine-driven Ic-BLs lacking GRB counterparts]{soderberg2010} improbable. 
As noted earlier, a CSM-interacting Ic-BL like 11qcj could still explain the properties of \src, but given the higher implied energy (and velocity) at the peak of the afterglow light curve (see Figure~\ref{fig:U_v}), we disfavor this explanation.
We can also consider the case of dynamical ejecta from neutron star mergers \citep{lee2020}.
Ejecta of $\sim10^{50}$ erg at speeds $\sim$0.1--0.5$c$ are expected from the simulations of neutron-star mergers, while the outflows from black hole-neutron-star mergers could be faster and may reach $\sim10^{52}$ erg \citep[e.g.,][]{rosswog2013}.
Hence, the dynamical ejecta explanation requires an exceptional circumstance.


\subsubsection{Electron power-law index, $p$}. Studies of GRB afterglows have shown that these ultra-relativistic transients generally have $p$ in the range $2.0\sim2.8$ (e.g., \citealt{fong2015,troja2019,makhathini2020}), while 
steeper values, $p\simeq 3.0$, are more common in mildly-/non-relativistic blast waves (with the exception of SN II; e.g., \citealt{weiler2002,chevalier_fransson2006,ho2019-2018cow,coppejans2020}, but also see \citealt{soderberg2010}). 
The index $p\simeq3$ derived for the late-time afterglow of \src\ may therefore be suggestive of a mildly-/non-relativistic transient, but this does not conclusively rule out an initially relativistic blastwave. 
For example, the index could change during the relativistic to non-relativistic transition.
Indeed, such a transition has been suggested for the TDE Swift J1644+57, which harbored an initially relativistic jet \citep{cendes2021}. 
In the case of GRB 030329, it is found that $p\simeq2.1-2.5$ during the relativistic and non-relativistic regimes, so the electron power-law index may not have changed appreciably \citep{frail2005,vanderhorst2008,mesler_pihlstrom2013}.

\subsubsection{Host galaxy and local environment.} 
The host galaxy of the transient is a blue compact dwarf, characterized by high specific star-formation rate and low metallicity.
It is similar to the hosts of Long GRBs, SN Ic-BL, SLSNe, and AT2018cow-like FBOTs.
\cite{lee2020} evaluated that $\sim$1\% of neutron star mergers/short GRBs may occur in such hosts.
However, the relatively dense CSM ($n\gtrsim10$~cm$^{-3}$) needed to explain the afterglow of \src\ demands further fine-tuning for the delay-time and makes such an explanation unlikely. 

\subsubsection{Rates.} In order to better understand the rates\footnote{This rate estimate is more precise and complementary to the volume-limited one presented by \cite{2018ApJ...866L..22L} since the discovery paper used only the first half of the VLASS Epoch 1 catalog and had a low completeness of the galaxy catalog \citep{2017ApJ...846...44O}.}
of radio transients like \src, we carried out a flux density-limited search for transients detected in FIRST \citep{white1997} and absent in the VLASS Epoch 1.0 \citep{gordon2021}.
We do not find any other radio transient (particularly, having luminosity larger than $\sim2\times10^{28}$ erg\,s$^{-1}$\,Hz$^{-1}$) that are $>$4 mJy at 1.4 GHz in FIRST and absent in VLASS (i.e., 3 GHz flux density $<$1 mJy; in this search we removed candidates that were nuclear and hence very likely to be AGN or TDE).
Considering this unique event found in the 10,000 deg$^2$ of FIRST, we calculate the corresponding Poisson 68\% confidence interval to be 0.4--2.4 events \citep{gehrels1986}.
This corresponds to $\mathfrak R(>4 {\rm mJy},1.4~{\rm GHz})=(4-24)\times10^{-5}$\,deg$^{-2}$ of the sky. 
Alternatively, assuming a timescale of $\sim$10 years above 4 mJy at 1.4 GHz and a peak luminosity of $\gtrsim10^{29}$ erg\,s$^{-1}$\,Hz$^{-1}$, we can estimate\footnote{More generally, we can also calculate an upper limit for luminous afterglows for all classes of transients (95\% confidence): $<3\times10^{-4}$ deg$^{-2}$ above 4 mJy at 1.4 GHz, or in terms of volumetric rate: $<$600\,Gpc$^{-3}$\,yr$^{-1}$.} 
the volumetric rate: 40--240\,Gpc$^{-3}$\,yr$^{-1}$ (68\% confidence interval\footnote{The 95\% confidence interval is 5--470\,Gpc$^{-3}$\,yr$^{-1}$}; median rate is 100 Gpc$^{-3}$ yr$^{-1}$).
%
%




This rate corresponds to $\sim$0.1\% of the rate of core-collapse SN \citep{taylor2014}, $\sim$1\% of SN Ic-BL \citep{kelly_krishner2012,graham_schady2016,ho2020-koala}, $\sim$10\% of SN Ic-BL \citep{graham_schady2016,ho2020-koala}, and comparable to the rates of SLSN-I ($\sim$30\,Gpc$^{-3}$\,yr$^{-1}$), AT2018cow-like events ($\sim$400\,Gpc$^{-3}$\,yr$^{-1}$) and estimates for LGRBs ($\sim$60\,Gpc$^{-3}$\,yr$^{-1}$)~ \citep{guetta2005,quimby2013,goldstein2016,ho2020-koala} in the local universe.
The rate is also consistent with that of binary neutron-star mergers, but in this case, and similarly for SLSN-I, it requires that $\sim$50\%--100\% of the mergers/explosions to produce luminous radio emission, which is not consistent with observations \cite[e.g.,][]{fong2015,horesh2016,2019ApJ...876L..10E,schroeder2020,makhathini2020,gwtc2-population}.
Therefore we can rule out SLSN-I and neutron star merger explanations for \src.

\subsubsection{Putting it all together.}
Typical radio SN II/Ib/Ic are ruled out based on energy and velocity.
Engine-driven SN Ib/c (2009bb-like) are disfavored on energetic grounds.
SN Ic-BL are disfavored based on the peak luminosity and timescale as well as the long-lived radio light curve.
Considering the energy and velocity of the blastwave, we believe that SN Ic-BL with CSM interaction (11qcj-like) is unlikely.
From the small sample of AT2018cow-like events, this class of transients seems unlikely on account of the shape of the afterglow light curve.
Transient rates suggest that SLSN-I and neutron star mergers are unlikely. 
On the other hand, the afterglow light curve, velocities, blastwave energy, host galaxy properties and transient rates are compatible with the LGRB class.
We therefore conclude that, in terms of previously-studied transients, the afterglow properties of \src\ are most consistent with those of LGRBs.
A further exotic explanation, e.g., involving a magnetar or stellar merger, may not be required (but see below for a short discussion on  the magnetar scenario).

\subsection{Similarity with LGRBs; inverse beaming fraction and jet opening angle}
We conclude based on the above arguments that an LGRB remains the most likely explanation \citep{2018ApJ...866L..22L,2019ApJ...876L..14M} for \src.
Under this premise, we can calculate the inverse beaming fraction ($f_b^{-1}\equiv(\theta_j^2/2)^{-1}$) for GRBs, where $\theta_j$ is the average jet half-opening angle.
Using the formalism of \cite{2002ApJ...576..923L}, we calculate this parameter as\footnote{We have ignored the parameter $\tau_i$, the time at which the radio source just becomes isotropic. This way we assign it the same value, 3 years, considered by \cite{2002ApJ...576..923L}. We note that the dependence on this parameter is weak ($f_b^{-1}\propto \tau_i^{7/20}$).},

\begin{eqnarray}
f_b^{-1} &&\simeq 
230 ~N^{-1} \left(\frac{\mathcal{R}}{1\,{\rm Gpc}^{-3}\,{\rm yr}^{-1}}\right)^{-1}\,
\left(\frac{n}{10\,{\rm cm}^{-3}}\right)^{-19/24}
\nonumber\\
&& \times \left(\frac{E}{10^{51}\,{\rm erg}}\right)^{-11/6}\,
\epsilon_{e,-1}^{-3/2} \epsilon_{B,-1}^{-9/8}
 \nonumber \\
&&\simeq 
300 ~N^{-1} \left(\frac{\mathcal{R}}{1\,{\rm Gpc}^{-3}\,{\rm yr}^{-1}}\right)^{-1} \epsilon_{e,-1}^{-3/16} \epsilon_{B,-1}^{3/16} 
\nonumber\\
&&\times \left(\frac{R}{0.5\,{\rm pc}}\right)^{1/48} \left(\frac{t}{26\,{\rm yr}}\right)^{-19/12}
\label{eq:fbm1}
\end{eqnarray}


\noindent where $N$ is the number of afterglows detected in our search
above the minimum flux density threshold of 4~mJy, $\mathcal{R}$ is the observed rate of GRBs in the local universe ($z=0$) beamed towards us, $n$ is the CSM density, $R$ is the measured source radius in pc, $t$ is the age of the transient around epoch 2019.5, and we have used equation~\ref{eq:E_ISM}.
We have used the notation, $q_x = (q/10^x)$ for the microphysical parameters as in \S\ref{sec:Ben}.
Using $N=10^{0\pm0.4}$, $\mathcal{R}=10^{0\pm0.3}$ \citep{Schmidt2001,Wanderman&Piran10,lien2014}, $\epsilon_{B}=10^{-1\pm0.5}$, t=$27\pm1$ and using the normalization values for the other parameters in equation~\ref{eq:fbm1} we find $f_b^{-1}\simeq280^{+700}_{-200}$, corresponding to an average jet half-opening angle of $<\theta_j> \simeq5^{+4}_{-2}$\,degrees (68\% confidence), consistent with previous estimates \citep{frail2001,2002ApJ...576..923L,guetta2005,2006ApJ...639..331G,goldstein2016}.
\subsection{Predictions for future radio surveys and future evolution of \src}
The rate $\mathfrak R(>4 {\rm mJy},1.4{\rm GHz})$ derived above suggests that a radio survey across a hemisphere with $\sim$few mJy sensitivity should be able to find $\sim$1 \src-like transient.
Such surveys are currently being executed with the VLA \citep[the VLASS;][]{2020PASP..132c5001L}, ASKAP \citep[e.g.][]{mcconnell2020} and an even deeper survey has been proposed for the MeerKAT \citep{santos2016}.
Therefore, we predict that at least a few \src-like transients will be discovered in the coming years.

Since LGRBs are accompanied by GRB-SNe, it is possible that the late-time radio luminosity of \src\ will enter a second rise phase as the slower-moving SN ejecta collides with the ambient CSM \citep{BarniolDuran&Giannios15,kathirgamaraju2016,Peters+19,Margalit&Piran20,Eftekhari+21}. 
The relatively close distance of \src\ combined with its old age make this source an opportune target for detecting such emission, which would present unique possibilities for probing additional physics of the explosion \citep{Margalit&Piran20}. 
We therefore recommend continued radio monitoring of \src.


\subsection{An alternative explanation for \src\ involving magnetar}

Finally, we consider an alternative (speculative) possibility that \src~arose from a supernova (that may or may not be associated with a LGRB) that gave birth to a long-lived central engine, such as a millisecond pulsar or magnetar.  
In particular, a long-lived magnetar could power a nebula of synchrotron radio emission \citep[e.g.,][]{maruse2016,margalit_metzger2018}, which would become visible in radio once the supernova ejecta shell becomes optically thin to free-free absorption.
Although the ejecta in the case of SLSNe may take decades or longer to become optically thin at GHz frequencies \citep{margalit2018}, inconsistent with the rapid early light curve decay of \src, this transition could happen sooner for an explosion with a low ejecta mass (e.g., similar to those inferred in FBOTs or ultra-stripped supernovae).  
One motivation for this scenario is the speculation that the [OIII] emission line observed at late times from \src~could arise from the nebular phase of an engine-powered supernova, as a result of UV/X-ray emission from the engine nebula being reprocessed by the ejecta shell and cooling through the emission line.
While detailed nebular-phase photoionization calculations are currently challenging, a preliminary examination of spectra produced using CLOUDY (\citealt{Ferland+13}; see methods in \citealt{margalit2018}) indicates that, for parameters typical of SLSN magnetars and their ejecta, strong [OIII] emission broadly consistent with \src\ is common at $\sim$20 yr post-explosion.
However, more detailed calculations of the nebular phase of pulsar/magnetar-powered supernovae would be required to confirm this possibility and its implications for the ejecta structure.

\section{Conclusions}\label{sec:conclusions}
Based on all the observational data and our analysis of \srcfull, we arrive at the following conclusions.
\begin{itemize}
    \item \src\ is an unprecedented (radio-discovered, luminous, decades-long) transient having a peak radio luminosity $>2.3\times10^{29}$\,erg\,s$^{-1}$\,Hz$^{-1}$\ at 1.4\,GHz and detectable radio emission $>26$ years post-explosion. 
    \item Average blastwave velocity is $>$44,000 km~s$^{-1}$ in 1993 (assuming the first radio detection epoch is $<$1000 d post-explosion). If the rise of the light curve is as rapid as the decline then initial velocity of the transient is $\gtrsim$0.2c.
    \item Average blastwave velocity is $\simeq$19,000 km~s$^{-1}$ in 2019 (last observing epoch; assuming $\sim$26 yr post-explosion).
    \item Age- and CSM density-independent estimate of the blastwave energy is $\sim 5 \times 10^{50}$\,erg (dependent on the microphysical parameters).
    \item Optical spectroscopic observations from 2019 reveal a broad [OIII]$\lambda$4959,5007 emission line. We find that collisional-excitation in compact star-forming region(s) within the host galaxy is the most conservative explanation, but we cannot completely rule out its association with the transient.
    A transient origin for broad line could suggest the presence of a magnetar.
    \item Host galaxy properties are suggestive of a massive star progenitor that is more likely to form in high specific star-formation rate or low metallicity environments, similar to those observed for LGRBs, SN Ic-BL, SLSN-I, and AT2018cow-like FBOTs. We are able to rule out SLSN-I (and neutron star merger ejecta scenario proposed by \citealt{lee2020}) based on rates and peak radio luminosity, and find that SN Ic-BL (not associated with GRBs) is very unlikely based on the energetics.
    \item The observed afterglow properties of \src\ are most consistent with those of LGRBs in terms of previously-studied transients. The afterglow light curve is especially similar to the late-time evolution of GRB 030329 if we assume that the first radio detection occurred $\sim$1000 d post-explosion.
    \item If \src\ is a LGRB afterglow then the inverse beaming fraction is  $f_b^{-1}\simeq280^{+700}_{-200}$, and accordingly the average jet half-opening angle is $<\theta_j> \simeq5^{+4}_{-2}$\,degrees (68\% confidence).
    \item The late-time radio light curve of \src\ may reveal the presence of a GRB-SN and continued radio monitoring of \src\ is therefore recommended.
    \item The rates of \src-like events, which we find to be $4-24\times10^{-5}$ deg$^{-2}$ or equivalently about 40--240\,Gpc$^{-3}$\,yr$^{-1}$, suggest that the VLA Sky Survey and surveys with the ASKAP and MeerKAT
    will find a few such events over the coming years.
\end{itemize}

\acknowledgments
We thank Chuck Steidel for insights into the origin of the broad emission line in the optical spectrum, and the anonymous referee for comments that helped improve the clarity of the paper.
KPM thanks Wenbin Lu for discussions on mass-loss history and off-axis jet, and Anna Ho for discussions on SSA analysis. 
KPM is a Jansky Fellow of the National Radio Astronomy Observatory.
The National Radio Astronomy Observatory is a facility of the National Science Foundation operated under cooperative agreement by Associated Universities, Inc.
The NANOGrav project receives support from National Science Foundation (NSF) Physics Frontiers Center award number 1430284.
CJL acknowledges support from the National Science Foundation Grant 2022546.
BM is supported by NASA through the NASA Hubble Fellowship grant \#HST-HF2-51412.001-A awarded by the Space Telescope Science Institute, which is operated by the Association of Universities for Research in Astronomy, Inc., for NASA, under contract NAS5-26555.
KPM and GH acknowledge support from the National Science Foundation Grant AST-1911199.
The Dunlap Institute is funded through an endowment established by the David Dunlap family and the University of Toronto. B.M.G. acknowledges the support of the Natural Sciences and Engineering Research Council of Canada (NSERC) through grant RGPIN-2015-05948, and of the Canada Research Chairs program. 
Part of this research was carried out at the Jet Propulsion Laboratory, California Institute of Technology, under a contract with the National Aeronautics and Space Administration.
The NANOGrav project receives support
from National Science Foundation (NSF) Physics Frontier
Center award number 1430284.
\vspace{5mm}
\facilities{VLA, VLBA}

\software{\hbox{AIPS}, version 31DEC19 \citep{vanMoorsel1996}; \hbox{CASA}, v.~5.5.0-77 \citep{2007ASPC..376..127M}}

\appendix
\section{Synchrotron Model}
\label{sec:Appendix_Synchrotron}

For completeness, we provide below a brief derivation of the expressions for synchrotron emission that we utilize in our analysis.
The optically-thin synchrotron luminosity is
\begin{equation}
    L_\nu = \frac{3e^3}{m_e c^2} \zeta_e N_e (p-1) \gamma(\nu)^{-(p-1)} B ,
\end{equation}
where $\gamma(\nu) = \left( 2\pi m_e c \nu / e B \right)^{1/2}$ is the Lorentz factor of electrons whose characteristic synchrotron emission frequency is $\sim \nu$, $B$ the magnetic field, $N_e$ the number of electrons swept up by the blast wave, and $\zeta_e$ the fraction of these electrons that are relativistic and contribute to the synchrotron luminosity.
Within the framework of the deep-Newtonian regime \citep{Sironi&Giannios13}, the latter is given by
\begin{equation}
    \zeta_{e,{\rm eff}} = \frac{p-2}{p-1} \frac{m_p}{m_e} \epsilon_e \frac{1}{2} \left(\frac{v}{c}\right)^2 ,
\end{equation}
where $v$ is the shock velocity, and we assume that a fraction $\epsilon_e$ of post-shock kinetic energy goes into accelerating the non-thermal electron population.

Taking the ambient density to be either a constant density ISM, or an $r^{-2}$ wind, 
\begin{equation}
    \rho = 
    \begin{cases}
    n m_p &, \,{\rm ISM}
    \\
    A r^{-2} &, \,{\rm wind}
    \end{cases}
    ,
\end{equation}
the number of swept-up electrons can be expressed as a function of the shock radius $R$ as
\begin{equation}
    N_e = 
    \begin{cases}
    \frac{4\pi}{3} n R^3 &, \,{\rm ISM}
    \\
    4\pi A R / m_p &, \,{\rm wind}
    \end{cases}
    .
\end{equation}
Finally, assuming post-shock magnetic field amplification with an efficiency $\epsilon_B$ relates the magnetic field to the shock velocity and radius,
\begin{equation}
    B = 
    \sqrt{16\pi \epsilon_B \rho v^2} =
    \begin{cases}
    \left( 16\pi \epsilon_B n m_p \right)^{1/2} v &, \,{\rm ISM}
    \\
    \left( 16\pi \epsilon_B A \right)^{1/2} {v}/{R} &, \,{\rm wind}
    \end{cases}
    .
\end{equation}

Combining the above equations, we find that the optically-thin synchrotron luminosity is
\begin{equation}
    L_\nu = 
    \begin{cases}
    \frac{2 \pi e^3 m_p}{m_e^2 c^4} \left( 16\pi m_p \right)^{\frac{p+1}{4}} \left(\frac{2\pi m_e c}{e}\right)^{-\frac{p-1}{2}} (p-2) \epsilon_e \epsilon_B^{\frac{p+1}{4}} n^{\frac{p+5}{4}} R^{3} v^{\frac{5+p}{2}} \nu^{-\frac{p-1}{2}} 
    &, \,{\rm ISM}
    \\
    \frac{6\pi e^3}{m_e^2 c^4} \left(16\pi\right)^{\frac{p+1}{4}} \left(\frac{2\pi m_e c}{e}\right)^{-\frac{p-1}{2}} (p-2) \epsilon_e \epsilon_B^{\frac{p+1}{4}} A^{\frac{p+5}{4}} R^{-\frac{p-1}{2}} v^{\frac{p+5}{2}} \nu^{-\frac{p-1}{2}}
    &, \,{\rm wind}
    \end{cases}
\end{equation}
For the case where $p=3$, we find quantitatively that
\begin{equation}
    \nu L_\nu \underset{p=3}{\approx}
    \begin{cases}
    4.8 \times 10^{33} \, {\rm erg \, s}^{-1} \, \epsilon_{e,-1} \epsilon_{B,-1} n_0^{2} \left(\frac{R}{0.5 \, {\rm pc}}\right)^7 \left(\frac{t}{26 \, {\rm yr}}\right)^{-4} \left(\frac{m}{0.4}\right)^4 \nu^{0} 
    &, \,{\rm ISM}
    \\
    8.9 \times 10^{33} \, {\rm erg \, s}^{-1} \, \epsilon_{e,-1} \epsilon_{B,-1} A_\star^{2} \left(\frac{R}{0.5 \, {\rm pc}}\right)^3 \left(\frac{t}{26 \, {\rm yr}}\right)^{-4} m^4 \nu^{0}
    &, \,{\rm wind}
    \end{cases}
\end{equation}
where above we have expressed the shock velocity as $v = m R/t$ (eq.~\ref{eq:v}).
This is the same as eq.~(\ref{eq:nuLnu}) in the main text, and that is used to infer source properties.

\bibliographystyle{aasjournal}
\bibliography{fasttrants.bib}

\end{document}